\documentclass{aa}  

\usepackage{txfonts}
\usepackage[pdfpagelabels=false,colorlinks=true,linkcolor=blue,citecolor=blue,filecolor=blue,urlcolor=blue]{hyperref}

\usepackage[utf8]{inputenc}
\usepackage{amsmath}
\usepackage{graphicx}
\usepackage{float}
\usepackage{placeins}

\usepackage{color}
\usepackage{xcolor} % for colors used in comments
\usepackage{soul}

\newcommand{\add}[1]{\textcolor{orange}{#1}}
\newcommand{\del}[1]{\st{#1}}
\renewcommand{\add}[1]{#1}
\renewcommand{\del}[1]{}

\newcommand{\equ}[1]{eq.~(\ref{eq:#1})}

\newcommand{\equnp}[1]{eq.~\ref{eq:#1}}
\newcommand{\se}[1]{\S\ref{sec:#1}}
\newcommand{\fig}[1]{Fig.~\ref{fig:#1}}

\newcommand{\be}{\begin{equation}}
\newcommand{\ee}{\end{equation}}
\newcommand{\ba}{\begin{align}}
\newcommand{\ea}{\end{align}}
\newcommand{\bad}{\begin{equation} \begin{aligned}}
\newcommand{\ead}{\end{aligned} \end{equation}}
\newcommand{\bea}{\begin{eqnarray}}
\newcommand{\eea}{\end{eqnarray}}
\def\ra{\rangle}
\def\la{\langle}

\newcommand{\avg}[1]{\left\langle #1 \right\rangle}        % for average

\newcommand{\bul}{$\bullet\ $}

\newcommand{\no}{\noindent}

\newcommand{\msun}{M_\odot}
\newcommand{\Msun}{M_\odot}

\newcommand{\Zsun}{Z_\odot}

\newcommand{\ifm}[1]{\relax\ifmmode#1\else$\mathsurround=0pt #1$\fi}
\newcommand{\kms}{\ifmmode\,{\rm km}\,{\rm s}^{-1}\else km$\,$s$^{-1}$\fi}

\newcommand{\Mpc}{\,{\rm Mpc}}
\newcommand{\kpc}{\,{\rm kpc}}
\newcommand{\pc}{\,{\rm pc}}
\newcommand{\cm}{\,{\rm cm}}
\newcommand{\Gyr}{\,{\rm Gyr}}

\newcommand{\Myr}{\,{\rm Myr}}

\newcommand{\yr}{\,{\rm yr}}
\newcommand{\erg}{\,{\rm erg}}

\newcommand{\cmc}{\,{\rm cm}^{-3}}

\newcommand{\ltsima}{$\; \buildrel < \over \sim \;$}
\newcommand{\lsim}{\lower.5ex\hbox{\ltsima}}
\newcommand{\gtsima}{$\; \buildrel > \over \sim \;$}
\newcommand{\gsim}{\lower.5ex\hbox{\gtsima}}

\def\Mv{M_{\rm v}}
\def\Mveleven{M_{{\rm v},11}}
\def\Mveight{M_{{\rm v},10.8}}

\def\Rv{R_{\rm v}}

\def\Mg{M_{\rm g}}
\def\Ms{M_{\rm s}}

\def\Re{R_{\rm e}}
\def\fg{f_{\rm g}}

\def\Sig1{\Sigma_1}

\def\Rd{R_{\rm disc}}

\def\fb{f_{\rm b}}

\def\Rs{R_{\rm str}}

\def\eps2{\epsilon_{-2}}

\def\tff{t_{\rm ff}}

\def\tcool{t_{\rm cool}}
\def\tvir{t_{\rm vir}}
\def\eps{\epsilon}

\def\ssim{\!\sim\!}
\def\seq{\!=\!}
\def\ssimeq{\!\simeq\!}
\def\sequiv{\!\equiv\!}
\def\sgt{\!>\!}
\def\slt{\!<\!}

\def\sgtrsim{\!\gtrsim\!}
\def\slesssim{\!\lesssim\!}
\def\sgeq{\!\geq\!}
\def\sleq{\!\leq\!}
\def\sgg{\!\gg\!}
\def\sll{\!\ll\!}
\def\sdash{\!-\!}
\def\stimes{\!\times\!}

\def\Rsh{R_{\rm sh}}

\def\Mdotv{\dot{M}_{\rm v}}

\def\tgen{t_{\rm gen}}

\def\sfr{{\rm SFR}}
\def\sfe{{\rm SFE}}
\def\sigs{\sigma_{\rm s}}
\def\Ngen{N_{\rm gen}}
\def\tgen{t_{\rm gen}}
\def\rcool{r_{\rm cool}}

\defcitealias{dekel23}{D23}
\defcitealias{behroozi19}{B19}
\defcitealias{ferrara23}{F23}

\begin{document}

\titlerunning{Predictions of FFB} 
\title{Feedback-free starbursts at cosmic dawn:}
\subtitle{Observable predictions for JWST}

\authorrunning{Li et al.}
\author{
   Zhaozhou Li$^1$ \fnmsep\thanks{E-mail: \href{mailto:zhaozhou.li@mail.huji.ac.il}{zhaozhou.li@mail.huji.ac.il}},
   Avishai Dekel$^{1,2}$ \fnmsep\thanks{E-mail: \href{mailto:dekel@huji.ac.il}{dekel@huji.ac.il}},
   Kartick C. Sarkar$^{1,3,4}$,
   Han Aung$^1$,
   Mauro Giavalisco$^5$,\\
   Nir Mandelker$^{1}$,
   \and
   Sandro Tacchella$^{6,7}$
}

\institute{
   $^1$ Center for Astrophysics and Planetary Science, Racah Institute of Physics, The Hebrew University, Jerusalem, 91904, Israel\\
   $^2$ Santa Cruz Institute for Particle Physics, University of California, Santa Cruz, CA 95064, USA\\
   $^3$ School of Physics and Astronomy, Tel Aviv University, Tel Aviv, 6997801, Israel\\
   $^4$ Department of Space, Planetary \& Astronomical Sciences and Engineering, Indian Institute of Technology Kanpur, 208016, India\\
   $^5$ University of Massachusetts Amherst, Amherst, MA 01003-9305, USA\\
   $^6$ Kavli Institute for Cosmology, University of Cambridge, Cambridge, CB3 0HA, UK\\
   $^7$ Cavendish Laboratory, University of Cambridge, Cambridge, CB3 0HE, UK
}

% \date{Received September 15, 1996; accepted March 16, 1997}
\AANum{A\&A}
% \def\jobname{aanda}

% \abstract{}{}{}{}{} 
% 5 {} token are mandatory

\abstract
{}
{We extend the analysis of a physical model within the standard cosmology that robustly predicts a high star-formation efficiency (SFE) in massive galaxies at cosmic dawn due to feedback-free starbursts (FFBs). 
This model implies an excess of bright galaxies at $z \sgtrsim 10$ compared to the standard models based on 
the low SFE at later epochs, an excess that is indicated by JWST observations.}
{Here we provide observable predictions of galaxy properties based on the analytic FFB scenario. These can be compared with simulations and JWST observations.
We use the model to approximate the SFE as a function of redshift and mass, 
assuming a maximum SFE of $\epsilon_{\rm max} \seq 0.2 \sdash 1$ in the FFB regime. 
From this, we derive the evolution of the galaxy mass and luminosity functions as well as the cosmological
evolution of stellar and star-formation densities. We then predict the star-formation history (SFH), 
galaxy sizes, outflows, gas fractions, metallicities, and dust attenuation, 
all as functions of mass and redshift in the FFB regime.}
{The major distinguishing feature of the model is the occurrence of FFBs above a mass threshold that declines with redshift. 
The luminosities and star formation rates in bright galaxies are predicted to be in excess of  
extrapolations of standard empirical models and standard cosmological simulations,
an excess that grows from $z \ssim 9$ to higher redshifts.
The FFB phase of $\sim\! 100\Myr$ is predicted to show a characteristic SFH that fluctuates on a timescale of $\sim\!10\Myr$.
The stellar systems are compact ($\Re \ssim 0.3\kpc$ at $z \ssim 10$ and declining with $z$).
The galactic gas consists of a steady wind driven by supernovae from earlier generations, 
with high outflow velocities (${\rm FWHM} \ssim 1400 \sdash 6700\kms$),
low gas fractions ($<\!0.1$), 
low metallicities ($\lesssim\!0.1\Zsun$), 
and low dust attenuation ($A_{\rm UV} \ssim 0.5$ at $z\ssim 10$ and declining with $z$).
We make tentative comparisons with current JWST observations
for initial insights, anticipating more complete and reliable datasets for detailed quantitative comparisons in the future.
The FFB predictions are also offered in digital form.}
{}

% up to 6 keywords

\maketitle
%
%-------------------------------------------------------------------

\section{Introduction}
\label{sec:intro}

%\defcitealias{dekel23}{D23}
%\defcitealias{behroozi19}{B19}

%Review of the FFB scenario  
It has been argued from first principles that within the standard cosmological paradigm, the star-formation efficiency (SFE) in massive galaxies at cosmic dawn is expected to be exceptionally high due to feedback-free starbursts (FFBs; \citealt{dekel23}, hereafter \citetalias{dekel23}).  SFE at cosmic dawn is predicted to be significantly higher than the low SFE that is valid at later epochs and lower masses, where star formation is assumed to be suppressed by stellar and supernova feedback and/or AGN feedback.
The SFE is defined here as the integrated fraction of the accreted gas onto a dark-matter halo that is converted to stars: 
\be
\epsilon = \frac{\Ms}{\fb\,\Mv} \, ,
\ee
where $\Ms$ is the galaxy stellar mass, $\Mv$ is the halo virial mass, and 
$\fb\seq \Omega_\mathrm{b}/\Omega_\mathrm{m}\seq 0.16$ is the cosmic baryon fraction. An alternative, instantaneous SFE can be defined by relating the star formation rate (SFR)
to the gas-accretion rate, $\epsilon'\seq {\rm SFR}/\fb\Mdotv$. While at lower redshifts, typically $\epsilon \ssim 0.1$ or lower (\citealt{behroozi19}, hereafter \citetalias{behroozi19}; see also e.g., \citealt{behroozi13,rodriguez17,Moster2018,Tacchella2013,Tacchella2018}),
it is predicted to be of order unity for the most massive galaxies at $z\ssim 10$ and above. 

% necessary conditions for FFB
The simple key idea behind the FFB scenario is that in star-forming clouds, there is a natural window that lasts about two million years between a starburst and the onset of effective stellar wind and supernova feedback. Thus, if the gas density in a star-forming cloud is above a threshold of $\sim\! 3\stimes 10^3\cmc$, the free-fall time is 
$\sim\! 1\Myr$ or shorter, allowing the star formation to proceed to near completion in a 
manner that is free of wind and supernova feedback.  \citetalias{dekel23} showed that several other necessary conditions for FFB are likely to be fulfilled under the specified characteristic high-density conditions. For example, if these starbursts occur in $\sim\! 10^{6\sdash7}\msun$ Jeans or Toomre clouds of sizes larger than $\sim\! 10\pc$, the surface density is sufficiently high for the $\sim\!1\Myr$ period to also be free of radiative feedback such as radiative pressure or photoionization \citep{menon23,grudic21}. By an interesting coincidence, at the FFB densities, the cooling time to star-forming temperatures of $\sim\!10$K is shorter than the free-fall time, as long as the metallicity is not negligible, allowing the short starbursts required for an effective time gap between the burst and the onset of feedback. 
A unique feature of FFB is that, at metallicities significantly lower than solar, the cooling time becomes shorter than the free-fall time only after the cloud has contracted to a density comparable to the FFB density, delaying star formation until the onset of FFB. 

Furthermore, it has been shown that the clouds undergoing FFB are shielded against SN-driven winds and UV radiation from post-FFB clusters of earlier generations, provided that the clouds are more massive than $\sim\! 10^4 \msun$. 
For a galaxy with a stellar mass of $\sim\! 10^{10} \msun$ in a halo of $\sim\! 10^{11}\msun$ at 
$z\ssim 10$, the FFB is predicted to occur in about $10^4$ globular-cluster-like clusters populating a compact, subkiloparsec-sized galaxy, divided into about ten generations during the global halo inflow time of $\sim\! 80\Myr$. The FFB scenario has been analyzed in two extreme configurations: clouds in a spherical collection of shells, or a disk, associated with low or moderate angular momentum in the cold streams that feed the galaxy, respectively, with qualitatively similar results.
While \citetalias{dekel23} suspected that a low metallicity in the star-forming clouds may help to diminish the potentially disruptive effects of stellar winds during the first free-fall times, we argue below that a subsolar metallicity is actually not a necessary condition for FFB.
Indeed, while the supernovae in post-FFB clusters generate metals and dust, the newly forming FFB clouds, which are shielded against the supernova-driven winds, are not significantly contaminated by metals or dust.

%Qualitative observable predictions.
\citetalias{dekel23} evaluated the regime in redshift and halo mass where FFBs with a high SFE are expected. These authors crudely predicted FFBs above the threshold line 
\be
\frac{M_\mathrm{v,ffb}}{10^{10.8}\msun} \sim \left( \frac{1+z}{10} 
\right)^{-6.2} \, . 
\label{eq:thresholds}
\ee
This should be modulated with quenching above a threshold halo mass of $\sim\! 10^{12}\msun$ or more because of suppression of the cold gas supply through the halos and/or AGN feedback \citep{Dekel06,dekel09,dekel19_gold}. 
The quenching mass, combined with \equ{thresholds},  only permits FFBs above a redshift of $z \ssim 6$, and then only below the quenching halo mass. 

%Purpose of this paper 
The purpose of this paper is to present predictions of the FFB scenario based on the analytic toy model of FFBs for various observable galaxy properties. These predictions are to be compared with current and especially new JWST observations, as well as future simulations. 
We compare the FFB predictions with a high-$z$ extrapolation of the ``standard'' empirical model, UniverseMachine 
\citepalias[UM,][]{behroozi19}, as well as with predictions from a ``standard" semi-analytic model (SAM,
\citealt{2019MNRAS.483.2983Y,2019MNRAS.490.2855Y,yung23}) \add{that was calibrated %only 
to match %
$z\sim 0$ observations}, 
which resemble current cosmological simulations where the conditions for FFB are not resolved by the subgrid models.
We add tentative, incomplete JWST observation data in certain figures
to provide an initial qualitative impression of the compatibility between the FFB scenario and observations. 
In order to help researchers perform comparisons to other models, simulations, and observations, 
we provide digital tables and codes for reproducing the predictions online.%
\footnote{\url{https://github.com/syrte/ffb_predict}}

%outline
The paper is organized as follows.
In \se{sfe} we describe how we represent the FFB scenario with a predicted SFE as a function of redshift and halo mass. % 2
In \se{lf} we present the predicted mass function and luminosity function at different redshifts,
    and the cosmological evolution of stellar density and SFR density. % 3
In \se{sfh} we address the predicted bursty star-formation history (SFH) and how it can be constrained by observations. % 4
In \se{wind} we introduce a steady wind model for the gas profile in an FFB galaxy, which is then used in the following 
               three sections. % 5
In \se{radius} we describe the predicted radii of galaxies at cosmic dawn. % 6
In \se{gas_frac} we estimate the gas fraction and metallicity in FFB galaxies. % 7
In \se{dust} we evaluate the expected dust attenuation in such galaxies. % 8
In \se{disc} we provide a discussion of certain issues. % 9
In \se{conc} we summarize our results and outline our conclusions. % 10

We adopt a flat $\Lambda$CDM cosmology hereafter  with parameter values 
$\Omega_\mathrm{m}\seq 0.3$, $\fb\equiv\Omega_\mathrm{b}/\Omega_\mathrm{m}\seq 0.16$,
$H_0\seq 70\kms\Mpc^{-1}$, $\sigma_8\seq 0.82$, and $n_\mathrm{s}\seq 0.95$,
which is close to the Planck cosmology (e.g., \citealt{Planck18}).

\begin{figure*} % 1 
\centering
\includegraphics[width=0.98\linewidth]{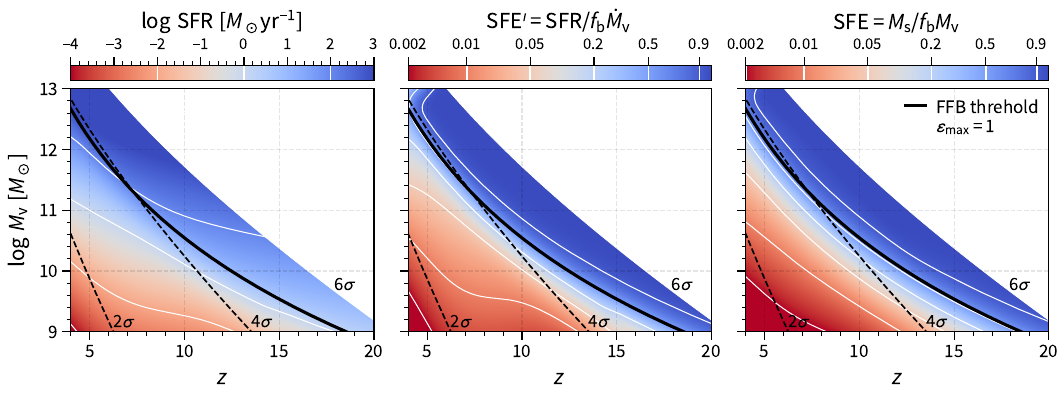}
\vspace{-5pt}
\caption{
Star formation rate and SFE 
adopted for the FFB model as a function of halo mass and redshift ($\Mv, z$).
{\bf Left:}
SFR.
{\bf Middle:}
Instantaneous SFE $\epsilon' \seq \sfr/(\fb\,\Mdotv)$.
{\bf Right:}
Integrated SFE $\epsilon \seq \Ms/(\fb \Mv)$.
The SFR and the instantaneous SFE are interpolated across the FFB threshold of \equ{thresholds} (solid black curve), 
from the low SFE based on the standard UM model \citepalias{behroozi19} well below the threshold,
to $\epsilon_{\rm max}\seq 1$ well above the threshold.
A smaller $\epsilon_{\rm max}$ gives similar results but with correspondingly lower values for FFB galaxies.
The colored area is limited to halos more abundant than 6$\sigma$ peaks,
with the halo masses of $2\sigma$ and $4\sigma$ peaks shown for reference (dashed lines).
The integrated and instantaneous SFEs are extremely similar.
}
%\vspace{-10pt}
\label{fig:sfr}
\end{figure*}

\begin{figure*} % 2 
\centering
\includegraphics[height=0.35\textwidth]{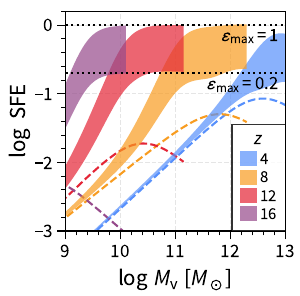}\hspace{1em}
\includegraphics[height=0.35\textwidth]{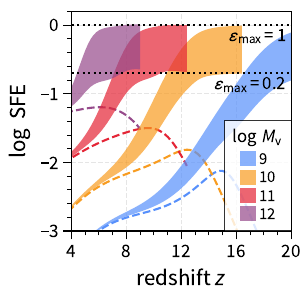}
\vspace{-3pt}
\caption{
Star formation efficiency adopted for the FFB model,
as in \fig{sfr}, but via curves of SFE versus $\Mv$ for a given $z$ (left) and SFE versus $z$ for a given $\Mv$ (right).
The SFE is made to coincide with the standard UM model \citepalias{behroozi19} (dashed lines) at low masses and redshifts and to approach $\epsilon_{\rm max}$ above the FFB thresholds of mass and redshift. 
The shaded areas refer to a maximum SFE that ranges from $\epsilon_{\rm max} \seq 0.2$ (bottom) to $1$ (top).
The FFB threshold is at roughly $0.5\,\epsilon_{\rm max}$.
The SFE does not reach $\eps_{\rm max}$ for $\Mv \sgeq 10^{12.5}\msun$ where SFR quenching is assumed. 
}
%\vspace{-10pt}
\label{fig:sfe}
\end{figure*}

%%%%%%%%%%%%%%%%%%%%%%%%%%%%%%%%%%%%%%%% 2
\section{Star-formation efficiency}
\label{sec:sfe}

As a basis for deriving observable predictions, we start with formulating the SFE
as a function of halo mass and redshift according to the FFB analytic model.

%===============================
\subsection{Star formation rate}
\label{sec:ffb_sfr}
% interpolation
The distinction between the FFB regime and the nonFFB regime is based on the threshold line $M_{\rm v,ffb}(z)$
given in \equ{thresholds}.
Well above this threshold, we assume that all the star-forming galaxies are FFB galaxies, while well below the threshold we assume that they are all nonFFB galaxies, with a smooth transition across the vicinity of the threshold line. 
We thus take the fraction of FFB galaxies among the star-forming galaxies to be
\be
f_{\rm ffb} (\Mv,z) = \mathcal{S} \left( \frac{\log[\Mv/M_{\rm v,ffb}(z)]}{\Delta_{\log M}} \right) \, ,
\label{eq:f_ffb_1}
\ee
where the smooth transition is arbitrarily approximated by a sigmoid function, $\mathcal{S}(x) \seq (1+\mathrm{e}^{-x})^{-1}$,  varying smoothly from zero to unity, and $\Delta_{\log M}$ is assumed quite arbitrarily to be 0.15 dex 
(with no qualitative effects on the results).

 % sfr in FFB
The mean SFR in an FFB galaxy is taken to be
\be
\la \sfr_{\rm ffb} \ra = \epsilon_{\rm max}\, \fb\, \Mdotv \, ,
\ee
where $\epsilon_{\rm max}$ is the maximum \emph{average} SFE valid in the FFB regime,
and $\Mdotv$ is the halo growth rate, which is approximately
\begin{equation}
    {\Mdotv}(\Mv, z)/\Mv \simeq s\, M_{\rm v, 12}^{\beta}\, (1 + z)^{2.5} \, ,
    \label{eq:Mdot}
\end{equation}
with $s=0.03{\mathrm{Gyr}}^{-1}$, $\beta \seq 0.14$ and $M_\mathrm{v,12}\equiv\Mv/10^{12}\msun$ 
\citep{neistein08a,dekel13}.

\add{We consider the global average maximum SFE $\epsilon_{\rm max}$ as a free parameter ranging from $0.2$ to unity. As demonstrated below, predictions with $\epsilon_{\rm max} \ssim 0.2$ appear to fit most of the current observational
estimates, while $\epsilon_{\rm max} \seq 1$ represents the theoretical upper limit for reference. 
FFB galaxies are predicted below to have a periodic bursty SFH (\se{sfh}), where the SFE fluctuates about the mean value $\epsilon_{\rm max}$, with peaks of SFE$\sim\! 1$ followed by periods of very low SFE. 
A value of $\epsilon$ below unity can be interpreted as the associated duty cycle,
given by the ratio of the characteristic duration of the peak ($\sim\!2\Myr$) to the typical duration of each cycle ($\sim\!10\Myr$).
}

We estimate below in \se{sfh} that the variation in halo accretion rate and the bursty SFH induce a scatter of $\sigma \ssim 0.3$ dex in the SFR.
Assuming $\mathrm{SFR}_\mathrm{ffb}$ to follow a lognormal distribution, the median $\log \mathrm{SFR}$ is then $\log\avg{\sfr_{\rm ffb}}-0.5\sigma^2\ln 10$.

 % sfr in nonFFB 
For nonFFB galaxies, we adopt the empirical model \textsc{UniverseMachine} (UM, \citetalias{behroozi19}).
The UM has been calibrated with pre-JWST observations, including stellar mass functions and quenched fractions (at $z\slt 4$), specific and cosmic SFRs ($z \slt 10$), UV luminosity functions ($4\slt z\slt 10$), and clustering ($z\ssim 0$), where each observable is reproduced over the available redshift range.
The SFR of star-forming and quenched galaxies, $\mathrm{SFR}_\mathrm{um,sf}$ and $\mathrm{SFR}_\mathrm{um,q}$, are modeled separately in UM as functions of halo mass and redshift, 
along with a quiescent fraction $f_\mathrm{q}(\Mv, z)$.
As a crude approximation, we extrapolate the UM fitting formulae from Appendix H of \citetalias{behroozi19} 
to nonFFB galaxies in the relevant mass and redshift ranges at cosmic dawn.

At a given $\Mv$ and $z$, the probability distribution of the SFR in star-forming galaxies
is assumed to be
\be
p (\mathrm{{SFR}}| \Mv, z) = f_\mathrm{{{ffb}}}\, p_\mathrm{ffb}(\mathrm{SFR})   
                            +(1 - f_\mathrm{{{ffb}}})\, p_\mathrm{um,sf}(\mathrm{SFR}) \, ,
\label{eq:sfr_ffb}
\ee
where the quantities on the right-hand side are all at a given $(\Mv, z)$. % clarified
As in \citetalias{behroozi19}, the SFR of different types, $\mathrm{SFR}_\mathrm{ffb}$ and $\mathrm{SFR}_\mathrm{um,sf}$, are assumed to follow separately lognormal distributions, 
$p_\mathrm{ffb}$ and $p_\mathrm{um,sf}$.
The mean SFR for a given halo mass is then $\la\sfr |\Mv, z\ra \seq \int \sfr\, p(\sfr|\Mv, z)\, \mathrm{d} \sfr$.

When referring to the whole galaxy population, the expression in \equ{sfr_ffb} should be multiplied by 
$(1-f_{\rm q})$, and an additional term, $f_{\rm q}\,p_{\rm um,q}(\mathrm{SFR})$, should refer to the quenched 
galaxies. The latter is assumed to dominate above a certain threshold quenching mass, 
where the SFR is suppressed by a hot CGM and/or the onset of effective AGN feedback \citep[e.g.,][]{dekel19_gold}.
While this quenching mass threshold is very uncertain in \citetalias{behroozi19} due to the lack of constraints at $z>4$, 
we find that the FFB predictions at $z \sgt 6$ are not sensitive to the quenching details,
as most high-$z$ galaxies are well below the quenching threshold and thus expected to be star-forming.

\begin{figure*} % 3
\centering
\includegraphics[width=1\textwidth]{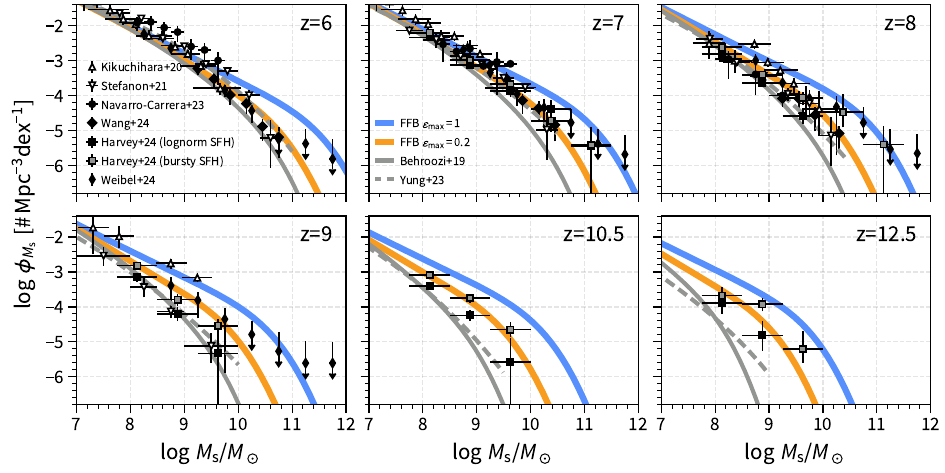}
\vspace{-10pt}
\caption{
Stellar mass function.
Shown are the predictions of the FFB model at $z\seq 6$--9,  $12.5$, and 16.
The blue and orange lines refer to $\epsilon_{\max} \seq 1$ and 0.2, respectively.
Shown for comparison are the extrapolated standard UM model \citepalias[solid gray line]{behroozi19}
and the results of a standard semi-analytic model \citep[solid dashed line]{yung23}. 
The FFB model predicts a greater abundance of massive galaxies, particularly evident from $\Ms \sgtrsim 10^9\msun$ at $z \ssim 9$.
The excess for lower masses also becomes larger at higher $z$.
\add{The open symbols show pre-JWST observational constraints \citep{Kikuchihara2020,Stefanon21} and the filled symbols refer to tentative JWST constraints 
(\citealt{Navarro-Carrera2024,Wang2024,Harvey2024,Weibel2024}).
The FFB predictions with $\epsilon_{\rm max} \ssim 0.2$ seem to be consistent with the observed mass function at the massive end, though the evidence for deviations of the data from the standard UM model is only marginal for this statistic, especially at $z\sim9$. 
This is compared to the stronger evidence provided below by other observables.
Indeed, we should note that SMF measurements suffer from large uncertainties in both observation and modeling.
For example, assuming a bursty SFH (as expected by the FFB scenario, \se{sfh}) rather than a smooth SFH during SED fitting results in an increased abundance of massive galaxies
(gray squares; \citealt{Harvey2024}).
}
}
% \vspace{-3pt}
\label{fig:stellar_MF}
\end{figure*}

%========================
\subsection{Stellar mass}
\label{sec:ffb_mstar}

\begin{figure*} % 4 
\centering
\includegraphics[width=0.74\textwidth]{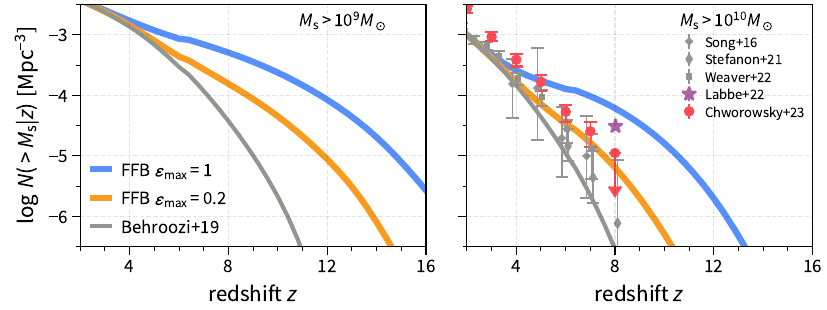}
\vspace{-2pt}
\caption{
Distribution of stellar masses.
Shown are the FFB predictions for the
number density of galaxies with stellar masses above $10^{9}\msun$ (left panel) or $10^{10}\msun$ (right panel).
The blue and orange lines refer to $\epsilon_{\max} \seq 1$ and 0.2, respectively.
The standard UM model \citepalias{behroozi19} is shown for comparison (gray line).
The FFB predictions are above the UM model for $z \sgeq 6$, and more so at higher redshifts.
Shown for tentative comparison are the observational measurements compiled by \citet{Chworowsky2023}, including JWST data (\citealt{labbe23,Chworowsky2023}) as well as earlier HST data 
\citep{Song2016,Stefanon21,Weaver2022}.
The JWST measurements are systematically higher than the previous measurements and they are above the standard 
UM model. The observations seem to favor the FFB prediction with $\epsilon_{\max} \sgtrsim 0.2$.
}
%\vspace{-10pt}
\label{fig:N_Mstar-z}
\end{figure*}

In the original UM model, the stellar mass was computed self-consistently by integrating the 
in situ SFR and the mass accreted through mergers. However, this is only workable with provided 
halo merger trees.
For simplicity, instead of resorting to detailed merger trees, we use here the median halo growth history. We compute the \textit{median} stellar mass by starting from the stellar mass of the UM model and adding the time-integrated difference between the mean SFR of the UM+FFB model ($\mathrm{SFR_{um+ffb}}$, \se{ffb_sfr}) 
and the mean SFR of the original UM model ($\mathrm{SFR_{um}}$), namely,%
\footnote{The median stellar mass comes from the median halo growth history, but when following the SFH we consider the mean SFR as an approximation. 
This is assuming that the stellar mass increment during a time interval $\Delta t$ can be approximated
by $\avg{{\rm SFR}}\Delta t$ when $\Delta t$ is much longer than the timescale for SFR fluctuations about its mean.}
\begin{equation}
    \Ms(\Mv, z) = M_\mathrm{s, um}+\int_{0}^{t(z)} 
        \left<\mathrm{SFR_{um+ffb}-SFR_{um}}\right> {\rm d}t \, ,
\label{eq:smhr}
\end{equation}
where the quantities on the right-hand side are at given $(\Mv,z)$ and
where $M_\mathrm{s, um}(\Mv, z)$ is taken from the median of the stellar-to-halo mass relation in the original UM (\citetalias{behroozi19}).%
\footnote{We adopted the fitting formulae in Appendix J with the first parameter set in Table J1 of \citetalias{behroozi19}.
We made a minor change to the original fitting formulae, 
as they become less accurate for high-$z$ massive galaxies compared to the raw UM catalog based on merger trees.
This discrepancy is largely fixed by manually forcing the slope ${\rm d}\log \Ms/{\rm d}\log \Mv\geq 0.3$.}
We note that the integration on the right-hand side is performed for the typical halo assembly histories $\Mv'(z'|\Mv,z)$.
Integrating \equ{Mdot}, approximating $1 + z \simeq (t / t_1)^{- 2 / 3}$ at $z \sgg 1$
with $t_1 \seq (2/3)\, \Omega_\mathrm{m}^{- 1 / 2} H_0^{- 1} \simeq 17.3\Gyr$ \citep{dekel13},
the mass-growth history of a halo of $\Mv$ at $z$ is
\be
M'_\mathrm{v, 12}(z' | \Mv, z) \simeq \left[ M_\mathrm{v, 12}^{- \beta} 
+ \frac{3}{2}\, s\, t_1\, \beta\, (z' - z) \right]^{- 1 / \beta} \label{eq:halohist} \, .
\ee

We assume $p (\log \Ms | \Mv)$ to follow a normal distribution.
While \citetalias{behroozi19} do not provide a fitting formula for the scatter,
we find that it can be approximated by
\be
\sigs = 0.1 + 0.3\,\mathcal{S}\left(\frac{(13.9-0.3 z) - \log \Mv}{0.2}\right) \, ,
\label{eq:sigs}
\ee
varying from $0.4$dex to 0.1dex from the low mass end to the high mass end.%
\footnote{We note that when $\epsilon_\mathrm{max} \seq 1$, introducing a scatter in stellar mass may 
bring the stellar mass to slightly above $\fb \Mv$ in the most massive halos. One should therefore consider the stellar mass 
function with $\epsilon_\mathrm{max} \seq 1$ as an upper limit at the massive end.
}

% \medskip % absorb the empty space
We show the SFR and the instantaneous and cumulative SFE ($\sfr/\fb\,\Mdotv$ and $\Ms/\fb\Mv$) as functions of halo mass and redshift in \fig{sfr}, 
and in an alternative way in \fig{sfe}, respectively.
The SFE is made to match the fiducial UM model \citepalias{behroozi19} in low mass and low redshift halos
and it approaches $\epsilon_{\rm max}$ in massive halos and high redshifts above the FFB threshold.

\add{The FFB model predicts a strong redshift dependence in the SFE and the stellar-to-halo mass ratio following 
the evolving FFB threshold halo mass (eq. \ref{eq:thresholds}). A cautionary note that should be mentioned is that certain pre-JWST data \citep{Stefanon21} indicate only a weaker evolution of the SFE with redshift at $z \slt 10$.}

%%%%%%%%%%%%%%%%%%%%%%%%%%%%%%%%%%%%%% 3
\section{Stellar masses, luminosities, and SFR density}
\label{sec:lf}

%=========================
\subsection{Stellar mass function}

The stellar mass function is given by
\be
\Phi (\log\Ms) = \int p (\log\Ms | \Mv)\, n (\Mv)\, \mathrm{d} \Mv \, ,
\label{eq:smf}
\ee
where the conditional probability distribution in the first term is assumed to be a lognormal distribution with the median given by \equ{smhr} and the standard deviation by \equ{sigs},
and $n(\Mv)$ is the halo mass function in the $\Lambda$CDM cosmology, adopted from \citet[with implementation by \citealt{Diemer2018}]{watson13} based on simulations that are reliable at $z \ssim 10$ and beyond (unlike most other similar work). 
The cumulative number density of galaxies with stellar mass above a threshold value is then
$N(>\!\Ms) \seq \int_{\log\Ms}^\infty \Phi (\log\Ms')\, \mathrm{d} \log\Ms'$.

 % figs stellar mass function
\fig{stellar_MF} shows the predicted stellar mass function at $z\seq 6$--9, $12.5$, and 16.
% \add{in comparison with pre-JWST and JWST measurements}.
As an alternative way to present the galaxy abundance, \fig{N_Mstar-z} addresses the stellar-mass distribution via the cumulative number density of galaxies 
with stellar mass above a threshold value, $\Ms \sgt 10^{9}\msun$ or $10^{10}\msun$, as a function of redshift.%
\footnote{The slight discontinuities at $z\seq 6$ in \fig{N_Mstar-z}, \ref{fig:UV_LF_20}, and \ref{fig:SFR-Mh-z} reflect the fact that  \citet{watson13} provide separate fits for the halo mass function below and above $z \seq 6$.}
The FFB predictions cover the range of $\epsilon_{\rm max} \seq 0.2 \sdash 1$ (areas between the orange and blue lines).
They are compared to the standard UM model \citepalias{behroozi19} and the standard SAM \citep{yung23}.
The number density of $10^{10}\Msun$ galaxies at $z \sgeq 8$ predicted by the FFB model with $\epsilon_\mathrm{max}\seq 0.2$ 
is 1dex higher than the standard models, and even more so at higher redshifts or when assuming a larger $\epsilon_\mathrm{max}$.
Also shown are the first observational results from JWST, tentatively indicating general consistency with the FFB model
for massive galaxies at high redshifts for $\epsilon_{\max} \sgtrsim 0.2$. 

\add{
The uncertainty in the estimates of stellar masses makes the comparison using the SMF as in \fig{stellar_MF}
particularly tentative, as indicated by the big scatter among the different datasets, especially at $z=9$.
This is because the present stellar libraries, photoionization templates, and the wavelength range covered by the photometry are likely inadequate. 
For example, the inclusion of the mid-infra bands (MIRI) can make a significant difference in the stellar mass inferred from the spectral energy distributions (SED) (\citealt{papovich23,Wang2024}).
We note that using a flexible bursty SFH in the SED fitting tends to yield more massive galaxies \citep{Harvey2024}, which makes the data agree better with the FFB model with 
$\epsilon_{\max}\seq 0.2$.
Despite the uncertainties concerning the SFHs of high-z galaxies \citep{Tacchella2022b}, certain studies suggested that allowing flexible nonparametric SFH may reproduce the true masses more reliably by accounting for outshining \citep{Gimenez-Arteaga2024,Narayanan2024}.
Theoretically, the FFB model indeed expects the high-$z$ massive galaxies to have bursty SHFs (\se{sfh}).
}

\begin{figure*} % 5
\centering
\includegraphics[width=1\textwidth]{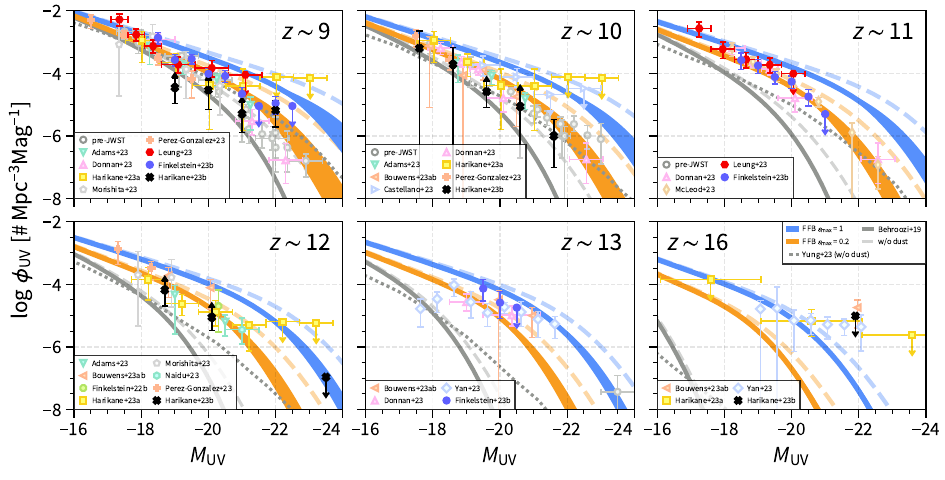}
\vspace{-15pt}
\caption{
Luminosity function.
Shown are the FFB predictions for the rest-frame UVLFs of galaxies at 
$z \seq 9$--13 and 16. 
The colored bands represent dust-corrected FFB predictions for $\epsilon_{\max} \seq 0.2$ (orange) or 1 (blue).
The dust correction assumes either the shell or the disk version of the FFB model at the bottom or top bounds of the shaded area, respectively.
The UVLFs uncorrected for dust are shown by dashed lines.
Shown for comparison is the fiducial UM model \citepalias{behroozi19} (gray), corrected for dust (solid) or uncorrected (dashed).
Also shown is the fiducial SAM simulation without dust attenuation \citep[][dotted gray]{yung23}.
The FFB predictions are well above the standard models, e.g., at $z \ssim 9$ for $M_{\rm UV} \sgeq -19$
and at $z \ssim 12$ for $M_{\rm UV} \sgeq -18$.
The open gray data points are pre-JWST constraints~(\citealt{McLeod2016,Oesch2018,Morishita2018,Stefanon2019,Bowler2020,Bouwens2021,Harikane2022_dropout,finkelstein22a}; only for $z \!\lesssim\! 10$).
The colored open symbols are measurements based on galaxies photometrically selected with JWST~(\citealt{Adams2023b, Bouwens2023b, Bouwens2023a, Castellano2022, Donnan2023, Finkelstein2022, Harikane2023, McLeod2023, Morishita2023, Naidu2022, Perez2023,Yan2023}; partly complied by \citealt{Shen23}).
In particular, bright-filled symbols are based on the JWST NGDEEP Epoch 1 \citep[red]{Leung2023}, 
the complete CEERS sample \citep[blue]{Finkelstein2023b},
and the JWST spectroscopically confirmed galaxies \citep[black]{Harikane2023-spec}.
At $z \ssim 11$, the FFB predictions seem to match the tentative bright-end JWST observations with 
$\epsilon_{\rm max} \ssim 0.2\sdash 0.5$.
We note that the photometric constraints at $z\ssim 16$ are highly uncertain \add{and possibly subject to redshift interlopers \citep{Harikane2023-spec}}. 
}
%\vspace{-10pt}
\label{fig:UV_LF}
\end{figure*}

\begin{figure*} % 6
\centering
\includegraphics[width=0.74\textwidth]{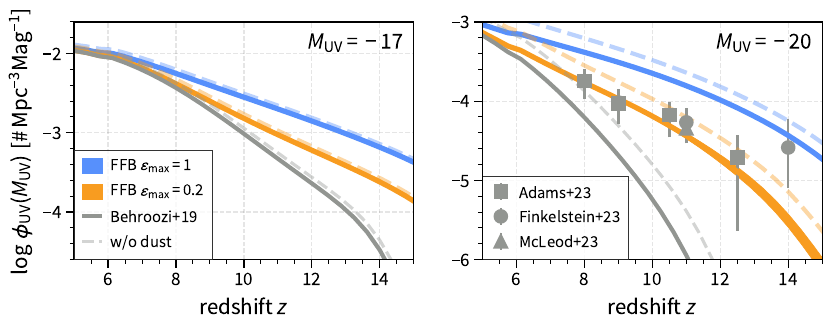}
\vspace{-5pt}
\caption{%
Evolution of the UV luminosity function at $M_\mathrm{UV} \seq -17$ (left) and $-20$ (right).
Similar to \fig{UV_LF}, the colored bands represent dust-corrected FFB predictions for $\epsilon_{\max} \seq 0.2$ (orange) or 1 (blue)
and the gray line shows the fiducial UM model \citepalias{behroozi19}. The UVLFs uncorrected for dust are shown by dashed lines.
The FFB predictions at $M_{\rm UV} \seq -20$ are higher than the UM model starting at $z \ssim 7$ and more so at higher redshifts. 
The open gray data points in the right panel are JWST measurements \citep{Adams2023b,Finkelstein2023b,McLeod2023}.
The tentative comparison indicates a fit with $\epsilon_{\max} \ssim 0.2$ at $z \seq 8 \sdash 12$ and possibly higher efficiencies at higher redshifts.
}
\label{fig:UV_LF_20}
\end{figure*}

\begin{figure*} % 7
\centering
\includegraphics[width=1\textwidth]{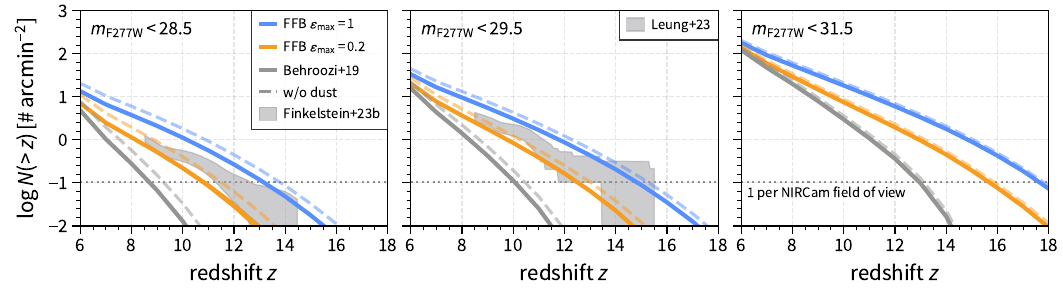}
\vspace{-13pt}
\caption{Cumulative surface number density of galaxies.
Shown is the surface number density per unit angular area at redshifts greater than $z$, for magnitude limits $m_\mathrm{F277W} \slt 28.5, 29.5$, and 31.5 in the three panels, respectively.
The FFB predictions are for $\epsilon_{\rm max} \seq 1.0$ (blue) and 0.2 (orange),
with (solid) or without (dashed) dust correction.
They are well above the standard UM predictions (gray).
JWST observations are shown in the left \add{and middle panels} (gray shaded area),
referring to the 68\% confidence interval for $z \sgt 8.5$ galaxies
from CEERS \citep{Finkelstein2023b} \add{and NGDEEP \citep{Leung2023} respectively}.
\add{The horizontal dotted line indicates the surface density corresponding to one galaxy per JWST NIRCam field of view ($9.7 \mathrm{arcmin}^2$).}
The tentative comparison indicates a fit with 
$\epsilon_{\max} \seq 0.2 \sdash 1$.
Future JWST observations (e.g., DESTINY proposed for Cycle 3) can possibly reach down to 
$m_\mathrm{F277W}\ssim 31.5$.
}
\label{fig:SurfaceDens}
\end{figure*}

\begin{figure*} % 8
\centering
\includegraphics[width=0.33\textwidth,trim={0.25cm 0 0 0},clip]
{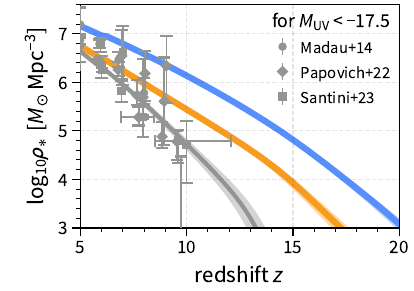}
\includegraphics[width=0.33\textwidth,trim={0.25cm 0 0 0},clip]
{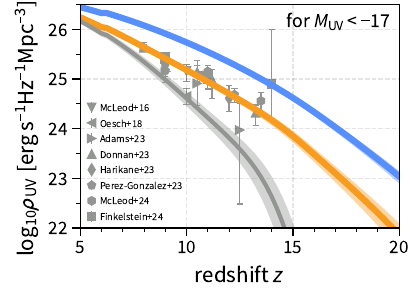}
\includegraphics[width=0.33\textwidth,trim={0.25cm 0 0 0},clip]
{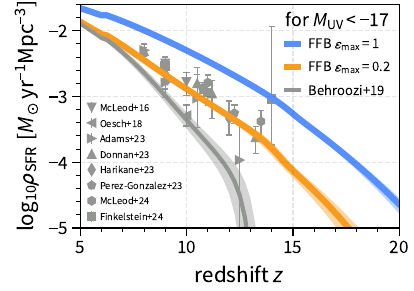}
\vspace{-13pt}
\caption{
Cosmological evolution of comoving densities.
{\bf Left:} Stellar mass.
{\bf Middle:} UV luminosity.
{\bf Right:} SFR.
The densities are computed based on galaxies brighter than $M_{\rm UV} = -17.5$ or $-17$ as indicated in the labels.
Shown are the FFB predictions for the bounding values of $\epsilon_{\rm max} \seq 0.2$ and $1$ (orange and blue respectively). 
Shown in comparison is the standard UM model \citepalias[][gray]{behroozi19}. 
The shaded bands represent rough uncertainty estimates that correspond to varying the magnitude limits by 0.5 mag up or down.
The FFB predictions are well above the UM model, starting with a small excess at $z \ssim 8$ and growing
to an order of magnitude at $z \ssim 12$.
Shown for tentative comparison are observational data (symbols).
The FFB model seems to match the tentative data for UV and SFR density with $\epsilon_{\rm max} \ssim 0.2$.
The references include measurements of  
stellar density \citep{Madau2014,papovich23,Santini2023},
UV density, and SFR density estimated from UV
(\citealt{McLeod2016, Oesch2018,adams23,Donnan2023,Harikane2023,Perez2023,McLeod2023,Finkelstein2023b};
see also e.g., \citealt{Coe2013,Ellis2013,Bouwens2020,Bouwens2023a,Bouwens2023b,Robertson2023}).
}
%\vspace{-10pt}
\label{fig:SFR-Mh-z}
\end{figure*}

%============================
\subsection{Luminosity function}

The UV luminosity function (UVLF) is derived from the halo mass function as
\be
\Phi (M_\mathrm{UV}) = \int p (M_\mathrm{UV}| \Mv)\, n (\Mv)\, \mathrm{d} \Mv \, ,
\label{eq:uvlf}
\ee
where $M_\mathrm{UV}$ is the absolute UV magnitude (in the AB magnitude system).
Here $p(M_\mathrm{UV}| \Mv)$, the probability distribution of the UV magnitude at a given halo mass and redshift, is assumed to be a Gaussian distribution. 
For the median, we adopt a simplified relation between the UV magnitude and the stellar mass \add{obtained from stellar population synthesis} in \citet[fig.~7]{yung23},
\footnote{\citet{yung23} showed that this median relation has a weak redshift dependence
\add{(also cf. \citealt{Stefanon21}, fig.~6)}, which we ignore here for simplicity.
We note that an alternative estimate of $M_{\rm UV}$ can be obtained by using the median SFH.}
\be
     \mathrm{median}\{M_\mathrm{UV}| \Mv \}  = -2.3 \log (\Ms / 10^9\msun) - 20.5 \, ,
\label{eq:muv-ms}
\ee
where $\Ms$ is taken from the median stellar-to-halo mass relation in \equ{smhr},
taking into account the enhanced luminosity in the FFB regime. % clarified
The variance of $\{M_\mathrm{UV}|\Mv\}$ is assumed to be $\sigma^2_{M_\mathrm{UV}}\seq 0.3^2+(2.3\sigs)^2$,  
combining an intrinsic scatter of 0.3 mag in $\{M_{\rm UV} \vert \Ms \}$ \citep[cf.][fig.~7]{yung23}
and an uncertainty propagated from the scatter in the stellar-to-halo mass relation (\equnp{sigs}), $2.3\sigs$,
where the factor 2.3 comes from \equ{muv-ms}.
We note that the high-luminosity end of the UVLF depends on the assumed variance \citep{Shen23}.

The UV magnitudes can be corrected for dust attenuation $A_\mathrm{UV}(\Mv,z)$, 
by replacing $\mathrm{median}\{M_\mathrm{UV}\}$ with $\mathrm{median}\{M_\mathrm{UV}\}+A_\mathrm{UV}$ in \equ{uvlf}.
The attenuation of FFB and nonFFB galaxies are incorporated separately,
using attenuation models from \se{dust} and \citetalias{behroozi19} (eq 23), respectively.
% \zzc{details:
% $p (M_\mathrm{UV}| \Mv)=f_\mathrm{ffb}p(M_\mathrm{UV}-A_\mathrm{UV, ffb}| \Mv) + (1-f_\mathrm{ffb})p(M_\mathrm{UV}-A_\mathrm{UV, um}| \Mv)$}

\fig{UV_LF} shows the FFB predicted UV luminosity function at $z=9$--13 and $16$, for $\epsilon_{\rm max} \seq 0.2$ and $1$.
Shown are both the uncorrected LF and the LF corrected for dust attenuation based on \se{dust}.
The LF from the FFB model is compared to the LF based on the standard UM model \citepalias{behroozi19}, with and without their correction for dust.
It is also compared to the LF in the standard SAM \citep{yung23} with no correction for dust.
The FFB LF at the bright end is significantly higher than the standard predictions, more so at higher redshifts.
For example, at $z \ssim 12$ the excess is higher by more than an order of magnitude
already at $M_{\rm UV} \ssim -20$.
Current observational results from pre-JWST and from JWST are displayed on top of the theory predictions.
While the scatter in the observational estimates is still large, 
the FFB model seems to provide a qualitative fit to the data,
with an SFE that increases with redshift, e.g., indicating $\epsilon_{\rm max} \seq 0.2 \sdash 0.5$ at $z \ssim 11$.

To highlight the excess of the luminosity function at high redshifts,
\fig{UV_LF_20} shows the redshift evolution of the UVLF at a given absolute magnitude, $M_\mathrm{UV}\seq -17$ or $-20$.
For example, at $M_{\rm UV} \ssim -20$, the FFB LF is higher than the UM model
by more than an order of magnitude already at $z \ssim 10$, and more so at higher redshifts.
Current observational results from JWST are displayed on top.
The fit of the FFB model to the data indicates $\epsilon_{\rm max} \ssim 0.2$ 
at $z \seq 8 \sdash 12$ and possibly higher efficiencies at higher redshifts.

\fig{SurfaceDens} shows the FFB-predicted surface number density of galaxies above a given redshift for different magnitude limits.
In order to compute this quantity, we converted the absolute magnitude $M_\mathrm{UV}$ to the JWST apparent band $m_\mathrm{F277W}$ using the luminosity distance and the K-correction.
A linear relation is assumed between the rest-frame $M_\mathrm{UV}$ and the observed-frame IR $m_\mathrm{F277W}$, whose slope and intercept are computed 
by fitting to a sample of galaxies in a semi-analytic model simulation
\citep{2019MNRAS.483.2983Y,2019MNRAS.490.2855Y},
in which spectral energy distributions (SEDs) were constructed 
based on the star-formation histories.
A tentative comparison to observational results from the complete JWST CEERS sample \citep{Finkelstein2023b} indicates consistency with the FFB model for $\epsilon_{\rm max} \seq 0.2 \sdash 1$.
\add{This is also consistent with the peak SFE estimated from the UVLF \citep[e.g.,][]{Sun2016, Sipple2024}.
}

%=========================================
\subsection{Evolution of the cosmic stellar mass and SFR densities}
\label{sec:sfr}

Given the stellar mass function from \equ{smf} and the luminosity function from \equ{uvlf}, 
the cosmic density of stellar mass and UV luminosity are derived as functions of redshift,
{\allowdisplaybreaks
\begin{align}
    \rho_{\rm s} (z) &= \int_{\la M_{\mathrm{UV}}|\Ms \ra < - 17.5} \Ms\, \Phi (\log\Ms |z)\, \mathrm{d} \log\Ms \, ,\\
    \rho_{\rm UV} (z) &= \int_{M_{\rm UV} < - 17.5} L_{\rm UV}\, \Phi (M_{\rm UV} |z)\, \mathrm{d} M_{\rm UV} \, ,
\end{align}}
where $L_\mathrm{UV}$ is the luminosity corresponding to $M_{\rm UV}$.
Similarly, the cosmic density of SFR can be estimated as
\be
\rho_{\sfr } (z) = \int_{\langle M_{\rm UV} | \Mv \rangle < - 17 }  \la\sfr |\Mv, z\ra\, n (\Mv | z)\, \mathrm{d} \Mv \, ,
\ee
where $\la\sfr |\Mv, z\ra$ is the mean SFR for a given halo mass
and $n (\Mv | z)$ is the halo mass function \citep{watson13}.
For comparison with observation, the above integrations are limited to galaxies brighter than $M_\mathrm{UV}\seq -17.5$ or $-17$.
In observation, $\rho_{\rm UV}$ is commonly used as a proxy for $\rho_{\rm SFR}$, through $\rho_{\rm SFR} \simeq \kappa_{\rm UV} \rho_{\rm UV}$
with a constant factor $\kappa_{\rm UV}\ssim 1.15\times 10^{-28}M_\odot \yr^{-1} \mathrm{erg^{-1}\, s\, Hz}$ that depends on the assumed initial mass function (IMF) and SFH \citep{Madau2014}.

\fig{SFR-Mh-z} shows the evolution with redshift of the average densities of stellar mass, UV luminosity, and SFR.
The FFB model predictions are shown for the bounding values of $\epsilon_{\rm max} \seq 0.2$ and $1$.
They are compared to the predictions of the standard UM model \citepalias{behroozi19},
showing a significant over-density for $z \sgeq 8$, which is growing with redshift.
Current observational results from JWST are displayed on top of the predictions, indicating general consistency with the FFB model, tentatively indicating $\epsilon_{\rm max} \ssim 0.2$.
We note that the comparison between different densities and corresponding observations may each lead to a somewhat different result. This is partly because the different densities as derived from observations may involve
different assumptions concerning quantities such as the IMF and the SFH, as well as the metallicity and dust obscuration, all affecting the conversion between SFR and UV emissivity.

%%%%%%%%%%%%%%%%%%%%%%%%%%%%%%%%%%%%%%%%%% 4
\section{Star formation history}
\label{sec:sfh}

\begin{figure*} % 9
\centering
\includegraphics[width=0.90\textwidth]{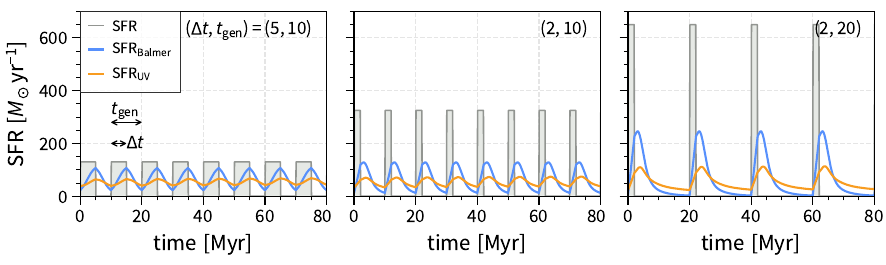}
\includegraphics[width=0.90\textwidth]{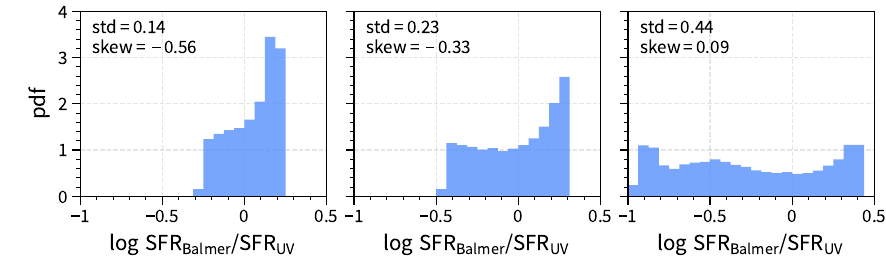}
\vspace{-5pt}
\caption{
Star formation history in an FFB galaxy at $z\ssim 10$.
The intrinsic SFH consists of several generations, separated by $\tgen\ssim 10\Myr$. 
Each generation consists of a peak of nearly simultaneous FFB starbursts that last for 
$\Delta t \seq 2\sdash5\Myr$, followed by a quiet period of gas accumulation until the onset of the following generation of bursts.
Shown are three cases where $\Delta t = 2, 5$ and $\tgen=10, 20 \Myr$.
{\bf Top:} Simplified representation of the periodic bursty SFH (gray histogram) and as deduced from two different SFR indicators, using a Balmer line (H$\alpha$, H$\beta$, etc.; blue) or FUV ($\sim\!1500 \AA$; orange), which probe the SFR at different time-scales.
{\bf Bottom:} Distribution of the ratio between the SFRs as indicated by Balmer lines and by FUV for galaxies with SFH in the corresponding top panels.
Quoted are the standard deviation and the skewness of the distribution, which can be used to constrain the characteristic $\Delta t$ and $\tgen$ by observations.
For a high duty cycle, $\Delta t/\tgen$, the distribution is dominated by a single skewed peak at a ratio larger than unity, while a long quiet period $\tgen-\Delta t$ broadens the distribution toward ratios smaller than unity.
This characteristic bursty SFH may bias the observed luminosities to even higher values at the bright end. 
}
\vspace{-3pt}
\label{fig:sfh}
\end{figure*}

As discussed in \S 7 of \citetalias{dekel23}, 
the main phase of star formation in an FFB galaxy is expected to last for a timescale comparable to the halo virial time $\tvir \ssim 80\Myr\,(1+z)_{10}^{-3/2}$, which is at $z \ssim 10$ in the same ball park as the time for accretion of most of the baryons (eqs.~31 and 32 there). 
This long phase is divided to $\Ngen\ssim 10$ consecutive generations, 
each of a period $\tgen \ssim \tvir/\Ngen$.
Each generation starts with a peak episode that consists of FFB starbursts in many clusters and consumes gas fast.
This peak episode lasts for a duration $\Delta t$ of a few cloud free-fall times ($\sim\!$ a few Myr) and is expected to be followed by a low-SFR period until the newly accumulated gas is sufficient for a new episode of fragmentation, bringing the onset of the next generation. 
The peak SFR is expected to be a factor $\tgen/\Delta t$ higher than the average SFR.
Based on the accretion rate in eq.~31 of \citetalias{dekel23}, 
the latter is the SFE ($\epsilon$) times the average baryonic accretion rate, 
\be
\la {\rm SFR} \ra \simeq 65\Msun\yr^{-1}\,\epsilon\,\Mveight^{1.14}\,(1+z)_{10}^{5/2}
\, .
\label{eq:SFR}
\ee

 % tgen
The number of generations for the two extreme toy models of shells and disks that were considered in \citetalias{dekel23} are 
\be
\Ngen \simeq
\begin{cases}
    7.1\,\Mveight^{1/3}\,(1+z)_{10}\,n_{3.5}^{-1/2} & \text{(shell)}\\
    10.6\,\lambda_{.025}^{-5/6}\,\Mveight^{-0.05}\,(1+z)_{10}^{-1/3} & \text{(disc)}
    \, .
\end{cases}
\label{eq:Ngen}
\ee
For disks this is based on the generation mass given in eq.~60 of \citetalias{dekel23}, assuming $Q\seq 0.67$,
and the total mass $\fb\,\Mv$. 
For shells the generation mass is given in eq.~46 of \citetalias{dekel23},
with $\Rs$ from eq.~35, assuming $T\seq 10^4$K and stream spin $\lambda_{\rm s}\seq 0.025$,
and with $\Rv$ from eq.~23 there.
At the threshold for FFB, \equ{thresholds}, with $n\seq 10^{3.5}\cmc$, this becomes
\be
\Ngen \simeq
\begin{cases}
    7.1\,(1+z)_{10}^{-1.07} & \text{(shell)}\\
    10.6\,\lambda_{.025}^{-5/6} & \text{(disc)}
    \, .
\end{cases}
\label{eq:Ngent}
\ee
Then, the period of each generation is
\be
\tgen \simeq \frac{\tvir}{\Ngen} \simeq 
\begin{cases}
    11.4\Myr\,(1+z)_{10}^{-0.43} & \text{(shell)}\\
    7.7\Myr\,(1+z)_{10}^{-1.5}\,\lambda_{.025}^{5/6} & \text{(disc)}
    \, .
\end{cases}
\label{eq:tgen}
\ee
At $z \ssim 10$, the generation period is thus expected to be on the order of $10\Myr$, so we consider periods in the range $\tgen \simeq 5-20\Myr$.

 % Delta t
The peak duration reflects the spread in starburst time among the FFB clusters of a given generation. A lower limit would thus be the FFB time $\sim\!1\Myr$.
An upper estimate could be the cloud-crushing time of a typical FFB cloud under the cumulative supernova wind from the clusters of earlier generations, which based on eq.~17 of \citetalias{dekel23} is $\sim\!5\Myr$ for a $10^6\msun$ cluster. 
We therefore consider peak durations in the range $\Delta t \simeq 2-5\Myr$.

The SFH, with the two free parameters $\tgen$ and $\Delta t$,
can be constrained by the distribution of the ratio of two different SFR indicators that are sensitive to different 
timescales, based on a sample of FFB galaxies.
Here we use the SFR based on common SFR indicators such as  H${\alpha}$ (or other Balmer lines) and FUV 
(1350--1750$\AA$), %
which are mostly sensitive to stars with ages of a few Myr and tens of Myr, respectively. 

The upper panels of \fig{sfh} show the toy model SFH for different choices of $\tgen$ and $\Delta t$ and the associated H${\alpha}$ and FUV signals as a function of time.  
We note that the SFH as indicated by any Balmer line,
such as H$\alpha$, H$\beta$, etc., is the same at first order \citep{Tacchella2022}.
The responses of the SFR using Balmer emission lines and FUV emission follow the approach in \citet{Caplar19}.
We predict the luminosity evolution for a Simple Stellar Population (SSP) using the Flexible Stellar Population Synthesis code (FSPS; \citealt{Conroy09, Conroy10}).\footnote{\url{https://github.com/cconroy20/fsps}}
We adopt the MILES stellar library and the MIST isochrones, with a metallicity of $0.1 Z_\odot$ for the stars and for the gas phase (without a qualitative effect on the results).

As can be seen in the upper panels of \fig{sfh},
the Balmer peak height is suppressed and its width is extended compared to the intrinsic SFR peak, and these two effects are stronger for the FUV peak.
The extension is because the photons are emitted from stars that live longer than the burst width.
The duration for the Balmer lines is shorter because they require H-ionizing photons ($\lambda \slt 912\AA$) which are produced by massive short-lived stars, while the FUV photons are also produced by less massive, older stars. 
The bottom panels of \fig{sfh} then show the corresponding distributions of the ratio $\log\,\mathrm{SFR_{Balmer}/SFR_{UV}}$, 
assuming that the galaxies are observed in a random phase of their periodic SFH.
The distribution is narrow at a high ratio above unity when the duty cycle $\Delta t/\tgen$ is high.
In all cases there is a peak at high $\mathrm{SFR_{Balmer}/SFR_{UV}}$, corresponding to the early times within a generation when H$\alpha$ is higher than FUV.
If $\Delta t /\tgen$ is sufficiently low, and especially when $\tgen$ is high, at late times within a generation H$\alpha$ drops to low values while FUV is still significant, leading to a second peak of the distribution at low $\mathrm{SFR_{Balmer}/SFR_{UV}}$ values.
As illustrated by the numbers in the lower panels of \fig{sfh},
one can use two parameters to quantify the shape of the distribution, e.g., the overall standard deviation ($\sigma$) and the skewness ($\gamma_1$).
For example, while $\sigma \seq 0.14$ and $\gamma_1 \seq -0.56$ for the case $(\Delta t,\tgen) \seq (5,10)\Myr$,
they are $\sigma \seq 0.44$ and $\gamma_1 \seq 0.09$ for $(2,20)\Myr$.

The SFH illustrated here is obviously oversimplified, and is meant for qualitative purposes.
For example, the stochasticity in gas accretion can introduce an additional scatter of $\sim\! 0.3$ dex \citep{dekel13}, which may dominate the variations in the SFH.
% \add{the variation of $\log{\rm SFR_{UV}}$ of the three SFHs above are 0.11, 0.23, 0.20 dex, respectively.}
More realistic predictions for the SFH variability will be obtained using cosmological simulations, semi-analytical or hydrodynamical. These will allow predictions for the temporal power spectrum of the SFH (e.g., \citealt{Caplar19,Iyer2020,Tacchella2020}) and the corresponding observable characteristics. 

We recall that the bursty SFH, which is naturally predicted for the FFB phase, is likely to contribute to the enhanced luminosity function at the bright end, where the mass function is dropping steeply, causing low-mass galaxies to preferably scatter to high luminosities 
\citep{Shen23,sun23a,sun23b}. 
This will add to the main effect of enhanced SFE by FFB. 

%%%%%%%%%%%%%%%%%%%%%%%%%%%%%%%%%%%%%%%%%%%%%%%% 5
\section{Steady wind and gas profile}
\label{sec:wind}

An important element of an FFB galaxy is the outflowing wind driven by the cumulative supernova ejecta from the earlier generations of star clusters within the galaxy while the outer shell is forming new FFB star clusters.
In \citetalias{dekel23} we crudely modeled this wind based on the mass in each generation of FFB, and used it 
to evaluate the condition for shielding of the new FFB clouds against crushing by this wind 
and for estimating the shell radius in balance between the forces exerted by the outflowing wind and the inflowing streams. 
Here, we model the wind somewhat differently as a steady wind that depends on the average SFR and the associated SFE and not explicitly on the generation mass,
thus simplifying the expressions and reducing the dependence on free parameters. 
This leads to a prediction of the expected observed line width as a function of the SFE,
and the steady-state gas density profile as functions of the SFE, SFR, and galaxy radius.  
We use this model in the following sections to re-evaluate the galaxy radius in the shell version of the FFB model, and to estimate the observable gas fraction, metallicity, and dust attenuation, all as functions of mass and redshift.

\add{In the following analysis, we assume a constant SFE with the mean value as determined by the halo mass and redshift.
As discussed in \se{sfh}, FFB galaxies are expected to have bursty star formation histories with fluctuating SFE,
which could influence the results presented below such as the estimates of the galaxy size and dust attenuation.
% , e.g., by biasing the luminosities high at the bright end where the halo mass function is steeply declining with mass.
For the purpose of the current qualitative estimates, we tentatively ignore these effects here and defer to future work
the incorporation of the effects associated with the fluctuating SFE.
}

%============================
\subsection{Steady wind}
\label{sec:steady_wind}

We consider the outflow generated by supernovae from a stellar system within a sphere of radius $R$, assuming here for simplicity a constant SFR during the $\sim\!100\Myr$ formation time of the galaxy. The overall rates of mass and energy outflows ($\dot{M}_{\rm sn}, \dot{E}_{\rm sn}$) increase over time as more stars explode in supernovae. 
The wind from each supernova is strong for about $20\Myr$ (see Figs.~1, 2 of \citetalias{dekel23}). 
After such a period, the supernova outflow saturates in a steady wind with outflow rates that are determined by the SFR. 
Based on \citet{starburst99}, assuming $Z \sim 0.1 Z_\odot$, these outflows are 
\be
\dot{M}_{\rm sn} \simeq 0.2\, {\rm SFR} \, , \quad
\dot{E}_{\rm sn} \simeq 7\times 10^{41} \erg\, \mathrm{s}^{-1} \, 
{\rm SFR}_{\Msun \yr^{-1}} \, .
\label{eq:SN_rates}
\ee

We assume that the supernova ejecta loads additional interstellar gas into the wind, such that the mass outflow rate is related to ${\sfr}$ via a mass-loading factor $\eta$,
\be
\dot{M}_{\rm out} = \eta\, {\rm SFR} \, .
\label{eq:eta}
\ee
The wind velocity is 
\be
V_{\rm w} \simeq \left( \frac{2\dot{E}_{\rm sn}}{\dot{M}_{\rm out}}  \right)^{1/2}
        = 1,490 \kms \, \eta^{-1/2} \, .
\label{eq:v_wind}
\ee

The mass-loading factor $\eta$ can be expressed in terms of the SFE $\epsilon$, defined here to be 
\be
\epsilon = \frac{{\rm SFR}}{\dot{M}_{\rm in}} \, , 
\label{eq:eps}
\ee
where $\dot{M}_{\rm in}$ is the gas inflow rate onto the galaxy.
Assuming that the inflowing gas that fails to be converted
to stars constitutes the diffuse medium to be loaded onto the supernova wind, we write
\be
\dot{M}_{\rm out} = \dot{M}_{\rm sn} + (1-\epsilon)\dot{M}_{\rm in} \, .
\label{eq:mout}
\ee
Using \equ{SN_rates}, \equ{eta}, \equ{eps} and \equ{mout}, we obtain
\be
\eta = 0.2+\frac{1-\epsilon}{\epsilon}   
     = \epsilon^{-1} - 0.8 \, .
\label{eq:eta_eps}
\ee

% line width
Based on \equ{v_wind} with \equ{eta_eps}, the wind velocity is predicted to be high and growing with the SFE, reflecting the associated dilution of the ISM,
\be
V_{\rm w} =
\begin{cases}
3,333\kms  &  \eps = 1 \\
\ \ \ \ 727\kms &  \eps =0.2 \, .
%\ \ \ \ 491\kms & \eps = 0.1 \, .
\end{cases}
\label{wind_eps}
\ee
The observed line width at the wings is expected to be roughly ${\rm FWHM} \ssimeq 2\, V_{\rm w}$,
and the line may be split to two peaks symmetric about the zero velocity, depending on the actual geometry of the inflow and outflow.
For high SFE in the FFB phase, where the ISM is rather dilute, these high supernova-driven velocities may be comparable to the velocities expected from AGN-driven winds \citep[e.g.,][]{kokorev23,Harikane2023agn}.
The FFB supernova-driven winds may be distinguishable by the associated lower metallicity (\se{metallicity} below) and lower ionization, in terms of line ratios in the BPT diagram or electron densities \citep[e.g.,][]{Kocevski2023,maiolino23}. We note, however, that the line ratios may be quite different than usual in nonFFB star-forming galaxies or AGN due to
excess emission from the supernova wind at $>\!70$ eV \citep{Sarkar2022}.
Being only a function of $\epsilon$, the observed line width at the wings
can be used to constrain the maximum SFE in the FFB phase \add{with $\eps \sgtrsim 0.2$}.

The potential effect of radiative losses on the wind is discussed in \se{cooling}, where we find that
they are expected to be ineffective at high SFE though they may slow down the wind for $\eps \slesssim 0.2$.
The possible effect of galactic radiative-driven winds is discussed in \se{disc_rad_wind}.

\begin{figure*} % 10
\centering
\includegraphics[height=0.33\textwidth]{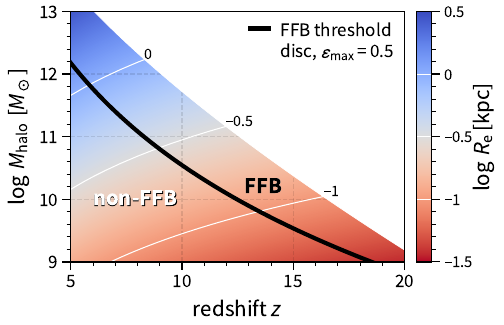}\quad
\includegraphics[height=0.33\textwidth]{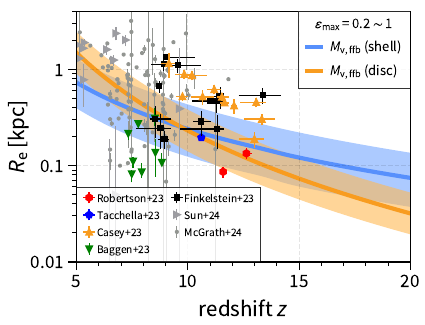}
\vspace{-5pt}
\caption{
Galaxy size.
{\bf Left:}
Galaxy size as a function of halo mass and redshift as predicted by FFB disk model assuming an interim value of $\epsilon_{\max}=0.5$.
{\bf Right:}
Predicted galaxy half-mass radius as a function of redshift for galaxies at the FFB threshold.
The estimates by the two limiting FFB models of shell (blue) and disk (orange), from \equ{Rsh_th} and \equ{Rd_th}, 
are pretty similar, predicting roughly $\Re \ssim 0.3\kpc\, (1+z)_{10}^{-2.5}$.
The color bands crudely indicate the uncertainty considering the scatter in the spin parameter. The dependence on $\epsilon_{\rm max}$ is weak.
Shown for comparison are tentative JWST observations
(\citealt{Baggen23,robertson23,tacchella23a,finkelstein23,Casey2023,Sun2024},
McGrath et al. in prep.).
The model provides a rough but reasonable estimate of the radius and effectively captures the observed decline of radius with redshift.
}
%\vspace{-10pt}
\label{fig:size_mz}
\end{figure*}

%==================================
\subsection{Gas density profile}
\label{sec:profile}

Given the outflow rates of mass and energy and the galaxy radius $R$,
the steady wind has a mass density profile with different asymptotic forms interior and exterior to $R$ \citep{Chevalier1985}, 
\be
\rho(r) = 
\begin{cases}
  0.3\,\left(\frac{\dot{M}_{\rm out}}{\dot{E}}\right)^{1/2}\, \frac{\dot{M}_{\rm out}}{R^2} 
  & r < R\\
  0.05\,\left(\frac{\dot{M}_{\rm out}}{\dot{E}}\right)^{1/2}\, \frac{\dot{M}_{\rm out}}{r^2}  
  & r > R
    \, .
\end{cases}
\label{eq:rho1}
\ee
The gas density within the galaxy radius $R$ is approximately constant.
Inserting the outflow rates from \equ{eta}, we obtain
\be
\rho(r) \simeq  
\begin{cases}
  1.1\times10^{-2}\,m_{\rm p}\,\cm^{-3}\, \eta^{3/2}\, {\rm SFR}_{\Msun \yr^{-1}}\, R_{\kpc}^{-2} 
  & r < R\\
  1.9\times 10^{-3}\,m_{\rm p}\,\cm^{-3}\, \eta^{3/2}\, {\rm SFR}_{\Msun \yr^{-1}}\, r_{\kpc}^{-2}
  & r > R
    \, .
\end{cases}
\label{eq:rho2}
\ee
The discontinuity at $r\seq R$ is due to a sharp transition from a subsonic wind within $R$ to a supersonic wind outside, which in practice is materialized by a continuous steep drop near $R$.

%%%%%%%%%%%%%%%%%%%%%%%%%%%%%%%%%%%%%%%%%%%%%%% 6
\section{Galaxy radius}
\label{sec:radius}

The typical radii of FFB galaxies can be predicted as a function of mass and redshift in the two extreme cases of disks and shells, based on the estimate of \citetalias{dekel23} for disks and a slightly revised new estimate based on a steady wind for shells.
Realistic galaxies might be mixtures of these two idealized cases.

%=============================
\subsection{Spherical shell model}

The outflow comes to a halt when it encounters the inflow and creates a strong shock
at a stalling radius $R_{\mathrm{sh}}$, which can be associated with the galaxy boundary.
The balance between the ram pressures of the outflow and inflow leads to
\be
\frac{R_{{\mathrm{sh}}}^2}{R_{{\mathrm{str}} }^2}
    = \frac{V_{\rm w} \dot{M}_{\rm out}}{4 V_{\rm in} \dot{M}_{{\mathrm{in}}}}
    = 1.13\, f(\eps)\, M_{v, 10.8}^{-1/3}\, (1 + z)_{10}^{-1 /2} \, ,
\label{eq:shell_size}
\ee
where the first equality is based on eq.~69 of \citetalias{dekel23} 
and the second equality is derived here based on the steady wind model, \se{wind}, replacing eq.~70 of \citetalias{dekel23}.
Here $R_{\rm str}$ is the effective radius of the streams,
$\dot{M}_{{\mathrm{in}}} \seq {{\mathrm{SFR}}}/{\epsilon}$ and $\dot{M}_{\rm out} \seq \eta\, {\mathrm{SFR}}$
are the inflow and outflow rates,
$V_{\rm w}$ is the wind speed (\equnp{v_wind}), 
$V_{\rm in}$ is the inflow stream speed, assumed to equal the virial velocity, eq.~24 in \citetalias{dekel23}.
\add{The factor $f(\epsilon)\sequiv\epsilon\sqrt{5\eta} \seq \sqrt{\epsilon(5-4\,\epsilon)}$ absorbs the dependence on $\epsilon$, with $f(\epsilon) \ssimeq 1$ for $\epsilon \sgtrsim 0.2$.}
Using the effective stream radius from eq.~35 of \citetalias{dekel23}, the shell radius becomes
\be
R_{\mathrm{sh}} =0.56 \kpc \times f(\eps)^{1/2}\, \lambda_{\rm s,.025}\, M_{v, 10.8}^{1/6}\, 
(1 + z)_{10}^{-3/4} \, ,
\label{eq:Rsh}
\ee
where $\lambda_{\rm s,.025}\equiv \lambda_{\rm s}/0.025$ is the contraction factor of the stream width 
with respect to the width of the underlying dark-matter filament \citep{mandelker18}.

At the FFB threshold $\Mveight\,(1+z)_{10}^{6.2}\seq 1$ from \equ{thresholds},
we obtain
\be
R_{{\mathrm{sh}}, {\mathrm{ffb}}} = 0.56\kpc \times f(\epsilon)^{1 / 2}\, \lambda_{s, .025}\, 
     (1 + z)_{10}^{- 1.78}\, .
\label{eq:Rsh_th}
\ee
The effective stellar radius may be very crudely assumed to be $\Re \simeq 0.5\,\Rsh$, 
allowing a contraction to virialization by a factor of two. %

We note that the shell radii obtained here using the steady wind model are only slightly different from those
obtained based on eq.~70 of \citetalias{dekel23}.
Both are determined specifically for the shell version of the FFB scenario. %

%=============================
\subsection{Disk model}

In the case of a disk, the typical gas disk radius is derived from the halo virial radius $\Rv$
and the contraction factor $\lambda$ to be
\be
R_{{\mathrm{d}}, {\mathrm{ffb}}} = \lambda \Rv = 0.31\kpc\, \lambda_{.025}\, \Mveight^{1/3}\, (1+z)_{10}^{-1}\, , 
\label{eq:Rd}
\ee
where $\Rv \seq 12.3\kpc\,\Mveight^{1/3}\,(1+z)_{10}^{-1}$ from eq.~23 of \citetalias{dekel23}.
We note that this disk radius is generic, not special for the FFB scenario.

At the FFB threshold $\Mveight\,(1+z)_{10}^{6.2} \seq 1$,
we obtain
\be
\Rd = 0.31\kpc\,\lambda_{.025}\,(1+z)_{10}^{-3.07} \, .
\label{eq:Rd_th}
\ee
This radius may reflect the stellar effective radius of the disk. %

One may crudely assume that the scatter in the size is dominated by the scatter in $\lambda$. If $\lambda$ reflects the halo spin parameter, it is well fitted in $\Lambda$CDM cosmological simulations by a lognormal distribution with a dispersion $\sigma_{\ln \Re} \ssim 0.5$ in the natural logarithm \citep{bullock01_j,danovich15,Yung23-GUREFT}, 
though this has not been reliably examined for $\sim\! 10^{11}\msun$ halos at $z \ssim 10$. % 

In summary, 
based on \equ{Rsh} and \equ{Rd}, we adopt for the galaxy radius (corresponding to $\sim\! 2R_{\rm e}$) in the two toy models of shell and disk
\be
R \simeq
\begin{cases}
    0.56\kpc\, f(\epsilon)^{1/2}\, \lambda_{\rm s,.025}\, \Mveight^{1/6}\, (1 + z)_{10}^{-3/4} \!\!\!\! & \text{(shell)}\\
    0.62\kpc\, \lambda_{.025}\, \Mveight^{1/3}\, (1+z)_{10}^{-1} & \text{(disc)}
    \, .
\end{cases}
\label{eq:R}
\ee
At the FFB threshold of \equ{thresholds}, the associated galaxy radius as a function of redshift is
\be
R \simeq
\begin{cases}
    0.56\kpc\, f(\epsilon)^{1/2}\, \lambda_{\rm s,.025}\, (1 + z)_{10}^{-1.78} \!\!\!\! & \text{(shell)}\\
    0.62\kpc\, \lambda_{.025}\, (1+z)_{10}^{-3.07} & \text{(disc)}
    \, .
\end{cases}
\label{eq:Rz}
\ee
While the radii in the two cases are comparable, the estimate for a shell is specific to the FFB model, while the estimate for a disk is rather generic to disk formation. %

The predicted galaxy half-mass radius, $R_{\rm e}$ (assumed to be $=\! 0.5\,R$ of \equnp{R}),
is shown as a function of halo mass and redshift in \fig{size_mz} for the toy models of shell and disk
with $\epsilon(\Mv,z)$ computed as explained in \se{ffb_sfr}.
Also shown is the half-mass radius of typical galaxies at the FFB threshold of \equ{thresholds}
as a function of $z$ (\equnp{Rz}), 
in comparison with a collection of tentative JWST observational measurements.
In order to focus on potential FFB galaxies, we display only the galaxies with an estimated stellar mass that is
larger than 10\% of the FFB threshold halo mass at the corresponding redshift.

%%%%%%%%%%%%%%%%%%%%%%%%%%%%%%%%%%%%%%%%%%%%%%%% 7
\section{Gas fraction and metalicity}
\label{sec:gas_frac}

\begin{figure} % 11
\centering
\includegraphics[width=0.85\columnwidth]{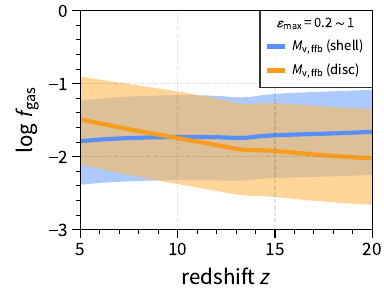}
\vspace{-3pt}
\caption{
Gas fraction.
Shown is $f_\mathrm{gas}\seq M_\mathrm{g}/(M_\mathrm{g}\!+\!\Ms)$ within $R\simeq 2\Re$
as a function of redshift at the FFB threshold, based on \equ{fg_z}, 
for $\epsilon_{\rm max} \seq 0.5$ in the shell and disk models.
The shaded bands indicate the uncertainty due to $\epsilon_{\max}$ ranging from 0.2 (top) to 1 (bottom). 
The gas fraction is rather low in the whole FFB regime, 
largely due to the low mass-loading factor associated with the high SFE.
}
%\vspace{-10pt}
\label{fig:gas_frac}
\end{figure}

%============================
\subsection{Gas fraction}
\label{sec:gas}

Integrating over the density profile from \equ{rho2}, the total gas mass in the galaxy is 
\be
\Mg = 1.16\times 10^6\msun\, \eta^{3/2}\, \sfr_{\msun\yr^{-1}}\, R_{1\kpc} \, .
\label{eq:Mg}
\ee
Inserting $\eta(\epsilon)$ from \equ{eta_eps}, SFR from \equ{SFR}, and $R$ from \equ{R},
we obtain the gas mass $\Mg$ for the disk and shell models as a function of $\epsilon$, $\Mv$ and $z$.
Dividing $\Mg$ by the stellar mass $\Ms \simeq \epsilon\,\fb\,\Mv$, we obtain the mass ratio
\be
\frac{M_\mathrm{g}}{\Ms} \!\simeq\!
\begin{cases}
0.61 \times 10^{-2}\, \eta_{1.2}^{7/4}\, \epsilon_{0.5}^{1/2}\, \lambda_{\rm s,.025}\, M_{v, 10.8}^{0.31}\, (1 + z)_{10}^{1.75} \!\!\!\!\! & \text{(shell)}\\
0.60 \times 10^{-2}\, \lambda_{.025}\, \eta_{1.2}^{3/2}\, \Mveight^{0.47}\, (1+z)_{10}^{3/2} & \text{(disc)} \, .
\label{eq:fg_Mz}
\end{cases}
\ee
Here $\epsilon_{0.5} \seq \epsilon/0.5$, and $\eta_{1.2}\equiv\eta/{1.2}$ refers to $\epsilon\seq 0.5$. 
For $\epsilon\seq 1$ or $0.2$, $\eta \seq 0.2$ or $4.2$ respectively, so
the ratio should be multiplied by $\sim 0.06$ or $6$ respectively.
At the FFB threshold given in \equ{thresholds}, we obtain as a function of redshift
\be
\frac{M_\mathrm{g}}{\Ms} \simeq
\begin{cases}
0.61 \times 10^{-2}\, \eta_{1.2}^{7/4}\, \epsilon_{0.5}^{1/2}\, \lambda_{\rm s,.025}\, (1 + z)_{10}^{- 0.15} \!\!\!& \text{(shell)}\\
0.60\times 10^{-2}\, \lambda_{.025}\, \eta_{1.2}^{3/2}\, (1+z)_{10}^{-1.43} \!\!\!& \text{(disc)} \, .
\label{eq:fg_z}
\end{cases}
\ee
When this ratio is small it is a proxy for the gas fraction, $f_\mathrm{gas} \seq M_\mathrm{g}/(M_\mathrm{g} \!+\!\Ms)$.

\fig{gas_frac} shows the gas fraction as a function of redshift at the FFB threshold
for the shell and disk models.
It is predicted to be rather low,
$\fg \slt 0.1$ for $\epsilon \sgt 0.2$.
A higher SFE leads to a smaller $\fg$ due to its dependence on $\eta$ in \equ{fg_z}, 
reflecting the smaller amount of residual gas left after star formation. %

%=========================
\subsection{Metallicity}
\label{sec:metallicity}

Based on their Fig.~1, showing the {\tt Starburst99} 
mechanical energy injection rate of stellar winds from a $10^6\Msun$ starburst, 
\citetalias{dekel23} expressed the worry that the winds may suppress the SFR if the metallicity in the star-forming clouds is $Z \sgeq 0.2\Zsun$. On the other hand, the simulations of \citet{Lancaster2021} indicated that mixing and cooling at the interface of the hot bubble and the cold ISM %
are likely to make the stellar winds ineffective for a few free-fall times even when the metallicity is as high as solar. Even if this conclusion is not valid under all circumstances, one can argue that the stellar winds should not be disruptive because, as we show here, the metallicity in the FFB star-forming clouds is expected to be low. 
The worry is that metals carried by supernova winds from the earlier generations of stars in the galaxy may contaminate the clouds in the outer shell or ring where the FFBs are expected to occur. This is unlikely for two reasons. First, the mixing of the metals in the clouds is expected to occur on a Kelvin-Helmholtz timescale. As evaluated in \S 4 of \citetalias{dekel23} in the context of the cloud shielding against the supernova-driven wind, this timescale is expected to be longer than the cloud free-fall time once the
clouds are more massive than $\sim\! 10^4\Msun$.  Second, we show below that even in the event of full mixing, the metallicity in the clouds is expected to remain relatively low, at the level of $Z \ssim 0.1\Zsun$.

\begin{figure} % 12
\centering
\includegraphics[width=0.85\columnwidth]{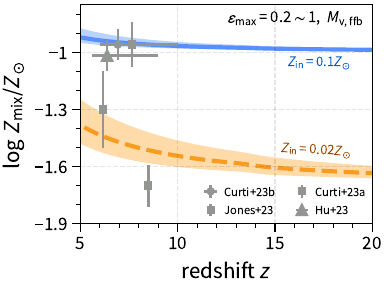}
\vspace{-3pt}
\caption{
Metallicity.
Shown is the upper limit for the metalicity in FFB star-forming clouds, assuming full mixing
between the outflowing metals $Z_\mathrm{sn} \seq 1\,Z_\odot$ and inflowing gas $Z_\mathrm{in} \seq 0.1\,Z_\odot$ (blue)
or $0.02\,Z_\odot$ (orange) in \equ{Z_mix}.
The metalicity is computed for halos at the FFB threshold mass.
The bottom and top bounds of the shaded areas represent $\epsilon_{\max}=0.2$ and unity, respectively.
Despite the mixing with supernova-enriched gas from earlier generations,
the metallicity remains low and close to $Z_\mathrm{in}$, at $Z \ssim 0.1\Zsun$ or below.
Shown for reference are tentative observations (\citealt{Curti2023a,Curti2023b,Jones2023,Hu2024}; also cf \citealt{Arellano-Córdova2022}), which suggest low but not negligible metalicities in the same ball park, though these galaxies are not necessarily in the FFB regime.
}
%\vspace{-10pt}
\label{fig:zmix}
\end{figure}

Assuming that the FFB clouds in the shell are limited to the areas of the shell that come into direct contact with the inflowing streams,
only a fraction of the outflowing metals participates in the mixing, $f_\Omega\equiv{\Omega_{\rm str}}/{4\pi}$,
where $\Omega_{\rm str}$ is the solid angle  corresponding to the streams. According to \equ{shell_size}, we have
\be
f_\Omega\equiv\frac{\Omega_{\rm str}}{4\pi} = \frac{\pi R_{\rm str}^2}{4\pi R_{\rm sh}^2} 
    = 0.22\, f(\epsilon)^{- 1} \Mveight^{1 / 3} (1 + z)_{10}^{1 /2},
\label{eq:Omega_str_1}
\ee
\add{where $f(\epsilon) \sequiv \sqrt{\epsilon(5-4\,\epsilon)} \ssimeq 1$  for $\epsilon \sgtrsim 0.2$.}
This gives in the FFB regime $f_\Omega \ssim 0.2$ rather insensitively to the value of $\epsilon$.

The mass of gas that outflows into the stream-fed regions of the shell (denoted by $'$) is
\be
\dot{M}'_{\rm out} = f_\Omega\, \dot{M}_{\rm out} = f_\Omega[\dot{M}_{\rm sn} + (1 - \epsilon) \dot{M}_{\rm in}],
\label{eq:Mdot_out}
\ee
(cf eq. \ref{eq:mout}).
Similarly, the outflow rate of metal mass into these areas is
\be
\dot{M}'_{{\mathrm{out}}, Z} = f_{\Omega}\, [\dot{M}_{{\mathrm{sn}}} Z_{{\mathrm{sn}}} + (1 - \epsilon) \dot{M}_{{\mathrm{in}}} Z_{{\mathrm{in}}}],
\label{eq:Mdot_out_Z_1}
\ee
The first term represents the metals ejected by the supernovae, with $\dot{M}_{\rm sn} \seq 0.2\,\sfr=0.2\epsilon \dot{M}_{\rm in}$ 
(see eq. \ref{eq:SN_rates}) and metalicity $Z_{\rm sn}$.
The second term represents the gas that came in by inflow, with metallicity $Z_{\rm in}$,
and has not turned into stars.

The outflowing metals and mass are assumed here to mix with the inflows, 
$\dot{M}_{{\rm in},Z}=\dot{M}_{{\rm in}} Z_{\rm in}$.
The mixed metallicity is then
\begin{align}
Z_{\rm mix\,} & = \frac{\dot{M}'_{{\rm out},Z}+ \dot{M}_{{\rm in},Z}}{\dot{M}'_{\rm out}+\dot{M}_{\rm in}}  
                = Z_{{\mathrm{in}}} + \frac{0.2\, \epsilon f_{\Omega}}{1 + (1 - 0.8 \epsilon) f_{\Omega}} (Z_{{\mathrm{sn}}} - Z_{{\mathrm{in}}}) \nonumber\\
              & < Z_{{\mathrm{in}}} + 0.2\, \epsilon f_{\Omega}\, Z_{{\mathrm{sn}}} \, .
\label{eq:Z_mix}
\end{align}
\fig{zmix} shows the mixed metallicity as a function of redshift at 
the FFB  threshold of \equ{thresholds}, assuming $Z_{\rm sn} \ssim 1\Zsun$ and $Z_{\rm in} \ssim 0.1 \Zsun$
for $\eps_{\rm max} \seq 0.2\sdash 1$.
The mixed metallicity is in the ball park of $Z_{\rm mix} \ssim 0.1\Zsun$ at all redshifts,
indicating that the enrichment by outflow is rather weak (because of the low $f_\Omega$ of penetrating cold streams) and quite insensitive
to redshift or $\epsilon_{\rm max}$.
This is a conservative upper limit for the metallicity in the FFB clouds.
The metallicity becomes even lower when assuming lower metallicity in the accreted gas, e.g., 
$Z_{\rm in} \seq 0.02\Zsun$.
Shown for comparison are tentative JWST observational measurements, which indicate low but non-negligible
metallicities in the ball park of the model predictions.
 
We conclude that as long as the observed gas metallicity is dominated by the gas in the FFB star-forming clouds, from which most of the UV luminosity is emitted, the predicted gas metallicity to be observed in a galaxy during its FFB phase is in the ball park of $Z \ssim 0.1\Zsun$. This is high enough for the cooling time to be
shorter than the free-fall time, as required for a burst at the FFB densities, and, in turn, it is low enough to guarantee that the stellar winds are not suppressing the FFB.

%%%%%%%%%%%%%%%%%%%%%%%%%%%%%%%%%%%%%%%%%%%%%%%% 8
\section{Dust attenuation}
\label{sec:dust}

%================================
\subsection{Attenuation at a given SFR}

Here, we use the global steady-wind model of an FFB galaxy to estimate its expected dust attenuation as a function of halo mass and redshift. The results are to be used to correct the predicted observable luminosities in the FFB regime, e.g., in the luminosity function.

 % N_H
Integrating over $\rho(r)$ in \equ{rho2} along a radial line of sight in the two regimes inside and outside the galaxy, we obtain the column density of Hydrogen atoms, 
\be
\begin{aligned}
N_H =& \int_0^R m_{\rm p}^{-1}\,\rho(r) \mathrm{d} r +
       \int_R^{\infty} m_{\rm p}^{-1}\,\rho(r) \mathrm{d} r\\  
   =& \,(3.44+0.58)\times 10^{19}\cm^{-2}\, \eta^{3/2}\, R_{\kpc}^{-1}\, {\rm SFR}_{\Msun \yr^{-1}} \, ,
\end{aligned}
\label{eq:NH}
\ee
where the first and second terms refer to integration over $r<R$ and $r>R$ respectively.

 % tau
Given $N_H$, the optical depth at 1500\AA\, is
\be
\tau_{1500} = N_H\, \sigma_{\rm ext,mw}\, f_{\rm d} \, ,
\label{eq:tau1}
\ee
where $\sigma_{\rm ext,mw} \ssimeq 10^{-21}\cm^2/{\rm H}$ is the corresponding extinction cross section of the dust per H nucleon in the Milky Way \citep{Weingartner2001}
and $f_{\rm d}$ is the dust-to-gas ratio in the outflow with respect to its value in the Milky Way.
The factor $f_{\rm d}$ in principle has contributions both from the supernovae 
($f_{\rm d,sn}$) and from the inflowing gas.
For low metallicity in the inflow, $Z_{\rm in} \ll Z_\odot$, relevant for FFB, the contribution from the inflow turns out to be negligible (see \se{dust_lowz}), such that
\be
f_{\rm d} \simeq 0.2\eta^{-1} f_{\rm d,sn} \, ,
\label{eq:fd_sn}
\ee
with $f_{\rm d,sn}\seq 5 \sdash 8 \ssimeq 6.5$ for $Z \ssim 0.1\,Z_\odot$ in stars \citep{Marassi2019}.
Inserting $N_H$ from \equ{NH} in \equ{tau1} we obtain
\be
\tau_{1500} \simeq (6.89+1.16) \times 10^{-3} \, f_{\rm d,sn}\, \eta^{1/2}\, R_{\kpc}^{-1}\, 
{\rm SFR}_{\Msun \yr^{-1}}\, .
\label{eq:tau_sfr}
\ee
The corresponding obscured fraction is determined from the optical depth as
$f_{\rm obsc} \seq 1-e^{-\tau}$.
The attenuation in UV is
\be
A_\mathrm{UV} = 2.5 \tau / \ln(10) = - 2.5 \log_{10} (1 - f_{\mathrm{obsc}}).
\label{eq:AUV}
\ee

 % f_obsc for shell and disc
Assuming that the new FFB star formation occurs mostly in a shell or a ring near $R$, such that the luminosity originates mostly there, the more relevant obscuration is by dust at $r>R$. In the shell toy model, one could very crudely assume that half the shell closer to the observer is obscured only by the dust in $r>R$, while the other half is obscured by the dust in $r>0$, such that the effective obscured fraction is the average of $f_{{\rm obsc},r>R}$ and $f_{{\rm obsc},r>0}$. 
In the case of a ring in the outskirts of a disk, for a line-of-sight that does not lie near the disk plane,
the effective obscuration may be very crudely estimated by $f_{{\rm obsc},r>R}$ alone.
For example, for $\epsilon \seq 0.5$, ${\rm SFR} \seq 65 \Msun\yr^{-1}$ and $R\seq 0.5\kpc$,
we estimate $f_{\rm obsc}$ in the ball park of $0.6 \sdash 0.8$.

\begin{figure} % 13
\centering
\includegraphics[width=1.03\columnwidth]{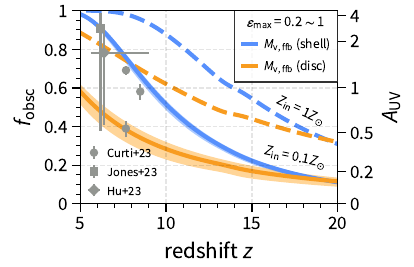}
\vspace{-12pt}
\caption{Obscured fraction $f_{\rm obsc}\seq 1-e^{-\tau}$ as a function of redshift $z$
along the threshold line for FFB, according to \equ{tau_th}. 
The corresponding extinction $A_\mathrm{UV}$ is shown on the right $y$-axis.
The toy models of the shell (blue) and disk (orange) are shown, with similar results.
The color bands show the uncertainty corresponding to $\epsilon_{\max}=0.2$ to 1
about $\epsilon_{\rm max} \seq 0.5$. %
We see a significant decline in obscuration with redshift.
The solid lines are for $Z_{\rm in} \seq 0.1$, typical in the FFB regime.
Shown for comparison (dashed line) are the results for 
$Z_{\rm in} \seq 1$, valid at later times, where the obscuration is stronger.
Shown for reference are tentative observations (\citealt{Curti2023a,Jones2023,Hu2024}; assuming a \citealt{Calzetti2000} attenuation law), which lie between the FFB predictions for shells and disks.
}
\vspace{-10pt}
\label{fig:fobsc_Mz}
\end{figure}

%==============================
\subsection{Attenuation in the FFB regime}

% % \tau (M,z)
We can further express $\tau$ as a function of $\epsilon$, $\Mv$ and $z$.
The average SFR, based on eq.~31 of \citetalias{dekel23}, is given by \equ{SFR}.
Using the galaxy radius 
from \equ{R} for the two toy models of shell and disk
adopting a contraction factor $\lambda_{.025}=1$,
and inserting these SFR and $R$ in \equ{tau_sfr}, we obtain
\be
\tau_{1500} \simeq
\begin{cases}
    (2.32+0.39)\, f(\epsilon)^{0.5} \, \Mveight^{0.97}\,(1+z)_{10}^{3.25} \hspace{-0.6em}& \text{(shell)}\\
    (2.12+0.36)\, f(\epsilon)\, \Mveight^{0.81}\,(1+z)_{10}^{3.5} \hspace{-0.6em}& \text{(disc)}
    \, ,
\end{cases}
\label{eq:tau_Mz}
\ee
where $f(\epsilon)\sequiv\epsilon\sqrt{5\eta} \seq \sqrt{\epsilon(5-4\,\epsilon)}$.
For the disk model we only consider $r \sgt R$, while for the shell model we adopt the average 
$f_{\rm obsc}$ in the ranges $r>0$ and $r>R$,
resulting in an effective prefactor of 0.36 for disks and 0.99 for shells in \equ{tau_Mz}.
Interestingly, the dependence of $\tau$ on the SFE $\epsilon$ is rather weak when $\epsilon \sgtrsim 0.2$, with $f(\epsilon) \seq 1,1.22,0.92$ for $\epsilon\seq 1,0.5,0.2$ respectively.
This is because the increase of SFR with $\epsilon$ is roughly balanced by the decrease of $\eta$ with $\epsilon$.

 % tau(z) at FFB threshold
At the FFB threshold as estimated in \equ{thresholds}, we obtain in \equ{tau_Mz}
\be
\tau_{1500} \simeq
\begin{cases}
    0.99\, f(\epsilon)^{0.5}\, (1+z)_{10}^{-2.78} & \text{(shell)}\\
    0.36\, f(\epsilon) \, (1+z)_{10}^{-1.50} & \text{(disc)}
    \, .
\end{cases}
\label{eq:tau_th}
\ee
\fig{fobsc_Mz} shows the obscured fraction as a function of redshift, based on \equ{tau_th} with $\epsilon_\mathrm{max} \seq 0.5$, valid for $Z_{\rm in} \ll 1$ (solid lines). The shell and disk toy models are compared. 
We see a significant decline of $f_{\rm obsc}$ with redshift for typical galaxies at the FFB threshold.
The tentative JWST observational measurements seem to lie between the FFB predictions for shells and disks,
though they do not necessarily correspond to FFB galaxies. 
The model predictions are also consistent with the low dust suggested by the very blue UV slopes for high-$z$ galaxies \citep{Topping2023,Morales2023}.

%=====================================
\subsection{Attenuation at low redshift and high metallicity}
\label{sec:dust_lowz}

% % Zin
For completeness,
we comment that at lower redshifts, away from the FFB regime, when $Z_{\rm in}$ (in solar units) is not much smaller than unity, the inflowing gas can make $f_{\rm d}$ grow significantly, and thus increase $\tau_{1500}$.
To see this,
define $\delta$ to be the dust-to-gas ratio in the different components.
Based on \citet{Remy-Ruyer2014}, $\delta$ in the inflowing gas is found empirically to be 
\be
\delta_{\rm in} = Z_{\rm in}^{1.6} \, \delta_{\rm mw} \, ,
\label{eq:Zin}
\ee
where $\delta_{\rm mw}$ is the ratio at solar metallicity, assumed to be valid in the ISM of the Milky Way.
By definition, $f_{\rm d} \seq \delta_{\rm out}/\delta_{\rm mw}$, 
where $\delta_{\rm out} \seq \dot{M}_{\rm dust} / \dot{M}_{\rm out}$.
For the dust we write 
\be
\dot{M}_{\rm dust} 
= \dot{M}_{\rm sn}\, \delta_{\rm sn} + (1-\epsilon)\,\dot{M}_{\rm in}\,\delta_{\rm in} 
\, .
\ee
Using $\delta_{\rm sn} \seq f_{\rm d,sn}\, \delta_{\rm mw}$ based on the definition of $f_{\rm d,sn}$, as well as \equ{eps} and \equ{Zin}, we get
\be
\dot{M}_{\rm dust}
= \dot{M}_{\rm sn}\, f_{\rm d,sn}\, \delta_{\rm mw} 
+ \epsilon^{-1}\,(1-\epsilon)\,{\rm SFR}\,\delta_{\rm mw}\, Z_{\rm in}^{1.6} \, .
\ee
Inserting $\dot{M}_{\rm dust}$ in 
$\delta_{\rm out} \seq \dot{M}_{\rm dust} / \dot{M}_{\rm out}$, 
and using \equ{SN_rates} and \equ{eta}, we finally obtain 
\be
f_{\rm d}  = \frac{\dot{M}_{\rm dust}}{\dot{M}_{\rm out}\delta_{\rm mw}}
= \eta^{-1}\, [0.2f_{\rm d,sn} + (\epsilon^{-1}-1)\, Z_{\rm in}^{1.6}] \, .
\label{eq:fd}
\ee
The second term is indeed negligible for $Z_{\rm in} \ssim 0.1 Z_\odot$, namely in FFB conditions, but it dominates if $Z_{\rm in} \ssim Z_\odot$ and $\epsilon \ll 1$, valid at low redshifts.
The obscured fraction for $Z_{\rm in}\seq Z_\odot$ is also shown in \fig{fobsc_Mz} for comparison.

%==============================
\subsection{The effect of cooling in the wind}
\label{sec:cooling}

The analysis so far neglected the possible effect of radiative cooling
of the wind within the galaxy, which could, in principle, weaken the outflow and enhance the dust attenuation  
if the effective cooling radius $\rcool$ is smaller than the galaxy radius $R$. 
Eq.~6 of \citet{Thompson2016} provides an estimate of the cooling radius in terms of the galaxy radius,
\be
\rcool \simeq 4\kpc\, \eta^{-2.92}\, R_{0.3}^{1.79}\, (0.3\,\sfr_{10})^{-0.79} \, ,
\label{eq:rcool}
\ee
where $\eta=\epsilon^{-1}-0.8$ is the mass loading factor (eq. \ref{eq:eta_eps}),
$R_{0.3} \seq R/0.3\kpc$
and $\sfr_{10}=\sfr/(10 \Msun \yr^{-1})$.
For a crude estimate, we insert the galaxy radius in the shell model from \equnp{Rsh} and the SFR from \equ{SFR} in \equ{rcool} to obtain
\be
\frac{\rcool}{R} \simeq 1,418 \times (5\epsilon^{-1} \sdash 4)^{- 2.72}\, \epsilon^{- 0.4}\, \Mveight^{- 0.77}\, 
(1 + z)_{10}^{- 2.57} \, .
\label{eq:rcool_R_Mz}
\ee
This ratio also scales with $\lambda_{\rm s,.025}^{0.79}$, which we assume here to equal unity. %
For the halos of the FFB threshold mass $M_{\rm h, 10.8} = (1 + z)_{10}^{- 6.2}$, this ratio becomes
\be
\frac{\rcool}{R} \simeq 1,418 \times (5\epsilon^{-1} \sdash 4)^{-2.72}\, \varepsilon^{-0.4}\, (1 + z)_{10}^{2.2}\, .
\label{eq:rcool_R_z}
\ee

We learn that for the high values of SFE and $z$ that are relevant for most of the FFB regime, the cooling radius is larger than the galaxy radius, so the correction for cooling is small. 
Cooling may become marginally effective at low $\epsilon$ or low $z$. 
For example, according to \equ{rcool_R_z}, at $(1+z) \seq 10$, $\rcool \ssim R$ for 
$\epsilon < 0.23$, namely cooling may be important at the low bound of the FFB regime.
At $z \sgtrsim 11$, on the other hand, $\rcool \sgt R$ even for $\epsilon \seq 0.2$, so cooling is unimportant in the whole range of SFE that is relevant for FFB.
In the nonFFB regime, where $\eps \sll 1$, $\rcool$ drops below $R$ due to the strong $\epsilon$ dependence in \equ{rcool_R_z}.
When $\rcool \sleq R$, the wind terminates at $r \slt R$ inside the galaxy, 
so the gas and dust are confined to radii smaller than $R$. In this case, 
if the star formation is mostly in an outer shell or a ring of radius $R$, 
the obscured fraction is expected to be in the range $0 - 0.5$ for ring or shell respectively, 
implying that less than half of the light from the star-forming region is obscured.
This is actually comparable to the result for a wind with no cooling.

There are several additional factors that can suppress the radiative cooling in the wind beyond the cooling rate that led to \equ{rcool}. 
First, \citet{Sarkar2022} showed that the cooling can be suppressed by nonequilibrium ionization in the wind and by excess radiation produced by the hot gas in the star-forming region. 
Second, entropy can be regenerated in the hot wind by bow shocks created at the interface between the wind and cold clouds \citep{Conroy2015,Nguyen2021,Fielding2021},
though it is possible that mixing of cold and hot gas near these clouds may overwhelm this 
increase in entropy. %
Based on all the above, we can conclude that the estimates of gas fraction, metallicity, and dust attenuation that neglected cooling in the wind are viable approximations in the FFB regime.

%%%%%%%%%%%%%%%%%%%%%%%%%%%%%%%%%%%% 9
\section{Discussion}
\label{sec:disc}

\subsection{Additional mechanisms for bright galaxy excess}
\label{sec:disc_mechanisms}

% mechanisms
Beyond the enhanced SFE due to FFBs presented in \citetalias{dekel23} and discussed here, 
there are several additional mechanisms that can potentially contribute to the excess of bright galaxies at cosmic dawn compared to extrapolations of the standard model and current simulations. Four different general types of additional mechanisms come to mind as follows.

\noindent$\bullet$
Scatter in the luminosity at a given halo mass, e.g., due to burstiness in the star-formation rate,
which preferentially scatters low-mass halos to higher masses at the massive end, where the halo mass function is declining steeply \citep{sun23b,yung23}. 

\noindent$\bullet$
Enhanced ratio of UV luminosity to stellar mass, e.g., by a top-heavy IMF, or by a nonstellar UV source such as AGN \citep{yung23,tacchella23b,grudic23}. 

\noindent$\bullet$
Reduction in dust attenuation at higher redshifts \citep{ferrara22,ferrara23}.

\noindent$\bullet$
Modifications of the standard $\Lambda$CDM cosmology that are associated with an enhancement in the abundance of massive halos, such as the introduction of Early Dark Energy \citep{klypin21} \add{or enhanced primordial power spectrum \citep{Parashari2023}}.

The enhanced SFE is robustly predicted by the physical model of FFB, which is naturally expected preferentially at the high densities and low metallicities present in high-mass halos at cosmic dawn. This mechanism alone can enhance the observed luminosities by an order of magnitude.
As shown in \se{sfh}, the FFB scenario naturally involves a characteristic bursty SFH at the bright end, which could bias the luminosity further to higher values.
Furthermore, as estimated in \se{dust}, the dust attenuation in the FFB regime is expected to be low and to decrease further with redshift, which also helps to make the galaxies brighter. 

\add{%
The consideration of a nonstandard IMF is currently under both theoretical and observational uncertainties, potentially either suppressing or enhancing the FFB mechanism and the abundance of early bright galaxies (see a more detailed discussion in \se{imf}).
}
The involvement of AGN in enhancing the UV luminosity is also conceivable but remains rather uncertain and yet to be explored; \add{tentative indications suggest that the contribution of AGN to the UV at cosmic dawn is subdominant \citep{scholtz23}.} 
Finally, turning to modifications of $\Lambda$CDM would necessarily involve ad hoc free parameters, potentially conflicting with the spirit of the simplicity principle of Occam's razor.

 % mass dependence
Overall, it is important to come up with observable features that could constrain the contribution of each mechanism to the excess of bright galaxies.
The main observable distinguishing feature of the FFB scenario is the preferred appearance of high SFE at high redshifts and possibly at high halo masses.
Such a mass dependence does not seem to emerge as a natural intrinsic feature if a top-heavy IMF is responsible for the bright excess, or if variations in dust attenuation are responsible for the redshift dependence of the excess (see \citealt{Finkelstein2023b}).  

\subsection{Initial mass function}
\label{sec:imf}

% IMF
\add{
The IMF at cosmic dawn is unknown. Given both the theoretical and observational uncertainties,
it could deviate from the standard low-redshift IMF either way, towards a top-heavy or a bottom-heavy distribution. 
Such deviations could either suppress or enhance the FFB nature of galaxy formation and the excess of early bright galaxies.}

\add{
On one hand, there are observational indications that the stellar populations in the cores of massive elliptical galaxies at low redshifts have a bottom-heavy IMF \citep{conroy12, vandokkum17, hallakoun21}.
If these are the descendants of the early FFB galaxies, in agreement with the average halo mass growth of three orders of magnitude from $z \ssim 10$ to the present, it would indicate that the FFB population had a bottom-heavy IMF. 
In agreement, simulations that include radiative and proto-stellar outflow feedback find that these processes tend to induce
an IMF with a lower characteristic stellar mass and an overall narrower mass range in high surface density environments \citep{tanvir22}, 
typical of the star-forming clouds at cosmic dawn.
When the IMF is bottom-heavy, the deficiency in OB stars would lead to suppressed supernova and stellar feedback, enhancing the FFB nature of galaxy formation at cosmic dawn 
even beyond the direct primary effects of high densities.
}

\add{
On the other hand, a top-heavy IMF may be expected at cosmic dawn due to the high cosmic microwave background (CMB) temperature or the high ISM pressure and density, as indicated in other simulations
\citep{grudic23,bate23,yung23}.
A top-heavy IMF is also indicated under certain assumptions from observations, e.g., in two nebular-dominated galaxies at $z \ssim 6-8$ \citep{cameron23} 
or by fitting the photometry of high-redshift galaxies with revised templates \citep{steinhardt23}. 
By enhancing the UV luminosity-to-mass ratio, a top-heavy IMF could contribute to an increased abundance of bright galaxies. However, typical simulations indicate only a limited factor of 2--3 in luminosity enhancement \citep{yung23}, 
which may not be sufficient by itself to explain the full excess of bright galaxies.   
Furthermore, the excessive radiation pressure associated with a top-heavy IMF will tend to suppress the SFE and thus act to reduce the excess of bright galaxies, leaving the net effect uncertain.
Simulations by Menon et al. (in prep.) find that a top-heavy IMF with a luminosity-to-mass ratio that is larger by an order of magnitude than the standard could reduce the SFE by a factor of two,
but the suppression is smaller at the low metallicities and dust content expected at cosmic dawn, where the radiative feedback in high-surface-density clouds is largely by IR radiation pressure on dust. This indicates that a top-heavy IMF and a high SFE can co-exist.}

\add{
In parallel, a top-heavy IMF can generate FFB in an alternative way, by accelerating the core collapse to black holes at the centers of the star-forming clusters (Dekel et al., in prep.). 
This process is driven by the inward migration of the massive stars due to dynamical friction, which would be naturally enhanced when more massive stars are present \citep{portegies02,rizzuto23}.
The core collapse of clusters is further sped up by the rotation and flattening of the clusters that form in galactic disks \citep{ceverino12}, via a process of gravo-gyro instability \citep{hachisu79,kamlah22}. 
In clouds where this core collapse occurs within a few Myr, it is expected to eliminate the massive stars before they can generate effective feedback at the end of their lifetimes, thereby leading to FFB.
}\add{
From the above discussion, we conclude that the IMF at cosmic dawn and its implications on FFB and the excess of bright galaxies remain largely uncertain, necessitating more thorough exploration.
}

%===========================
\subsection{Radiatively driven versus supernova-driven wind}
\label{sec:disc_rad_wind}

% Ferrara sSFR condition
The analysis of gas fraction, metallicity and dust in \se{gas_frac} and \se{dust}
is based on an assumed supernova-driven wind in steady state, as presented in \se{wind}. 
This wind is assumed to be driven by supernovae from earlier generations of stars, in their active supernovae phase, during most of the $\sim\! 100\Myr$ duration of the FFB galaxy formation phase. 
Here we comment on the validity of this supernova-driven wind in view of a possible global gas (and dust) ejection from the galaxy by super-Eddington radiative-driven winds, as proposed by \citet{ferrara23}.
Based on eq.~2 of \citet{ferrara23}, super-Eddington ejection is expected to occur once the specific SFR
in the galaxy is above a threshold,
\be
{\rm sSFR} > {\rm sSFR}_{\rm crit} \simeq 25 \Gyr^{-1}\, \left(\frac{100}{A}\right)\, 
\left(\frac{2}{f_{\rm bol}}\right)\, f_{\rm ts} \, .
\label{eq:sSFR_crit}
\ee
Here $A$ is the ratio of dust to Thomson scattering cross-section and $f_{\rm bol}$ is the bolometric correction with respect to the UV luminosity at $1500\AA$, both set to their fiducial values.
We introduce the correction factor $f_{\rm ts}$, the ratio of total to stellar mass within the galaxy, to account for the neglected contribution of the nonstellar mass to the Eddington luminosity. 
With a gas-to-stellar mass ratio of $\sim\!0.5$ and a moderate additional contribution of dark matter one can assume $f_{\rm ts} \ssim 2$. 
However, this factor may be balanced by a larger value of $A$, so we conservatively ignore it.

% z_crit is very high
\citet{ferrara23} assumes that the radiation dominates over the supernovae in driving the wind.
Tentatively adopting this hypothesis, we evaluate the redshift range of validity
of the condition in \equ{sSFR_crit}. Based on \citet{dekel13}, the analytic estimate in their eq.~7 and their simulation results, we approximate the sSFR by the specific accretion rate onto the halo,
\be
{\rm sSFR} \simeq 6.9\Gyr^{-1}\, (1+z)_{10}^{5/2}\, \Mveleven^{0.14}\, .
\label{eq:sSFR_z}
\ee
This redshift dependence is consistent with the findings of \citet{zhao09}
and it has been shown to be a better and more physical approximation than the crude estimate that is used by \citet{ferrara23} based on the ratio of halo mass to halo crossing time 
(see the difference between eqs.~31 and 32 of \citetalias{dekel23}).
Inserting the sSFR from \equ{sSFR_z} in \equ{sSFR_crit}, one expects super-Eddington ejection at
\be
1+z > 1+z_{\rm crit} \simeq 16.8\, \Mveleven^{-0.056}\, f_{\rm st}^{0.4}\, (A/100)^{-0.4}\, 
(f_{\rm bol}/2)^{-0.4} \, .
\label{eq:z_sSFR}
\ee
Assuming the fiducial values of the parameters, 
this implies a conservative lower limit of $z_{\rm crit} \ssimeq 16$ for halos of $\Mv \ssim 10^{11}\msun$.
If we adopt the weak mass dependence in \equ{sSFR_z}, and insert the declining 
$M_\mathrm{v, ffb}(z)$ at the FFB threshold from \equ{thresholds}, we can solve \equ{z_sSFR} for $z$
to obtain $z_{\rm crit} \ssimeq 22$.
The value of $z_{\rm crit} \seq 16$, and certainly $z_{\rm crit} \seq 22$, are beyond the redshift range of the current JWST observations in which the excess of bright galaxies is detected, making the radiative-driven global winds less relevant for the main FFB phase.

% SN vs radiative
At these extremely high redshifts when super-Eddington winds prevail, we return to 
the question of whether it is justified to neglect the SN-driven winds in the first place. \add{A quick comparison with the equivalent pressure produced by the SNe and radiation shows that these pressures are almost equivalent \hbox{\citep{TThompson2005, Murray2011}}. However, the SNe-driven pressure may be reduced due to radiative cooling such that the SNe pressure becomes much lower, and an energy-driven solution may not be viable for individual SNe. In such cases, the common wisdom is that a momentum-driven wind dominates.}
\add{Even in this case, it is found that} the wind-driving forces exerted by supernovae and by radiation from stars are comparable, at $\dot{p}/m_* \ssim 6\stimes 10^{-8}\cm\, {\rm s}^{-2}$ \citep{grudic20}, and both are comparable to the additional contribution from stellar winds.
% clustered SNe
We recall that the momenta from clustered sources, as expected in particular in the FFB compact clouds in compact galaxies at $z\ssim 10$, add up super-linearly for supernovae \citep{gentry17}, but linearly for radiative pressure, making the supernovae more important. \add{At significantly clustered star formation regions (as is the case for the high redshift galaxies under consideration), the wind becomes energy-driven again. Such highly clustered SNe can retain 20--$100\%$ of the SNe energy \hbox{\citep{Fielding2018}}, as estimated observationally in M82 \hbox{\citep{Strickland2009}}.}

% duty cycle
Furthermore, the \add{pre-blast wave} energy losses of SN bubbles due to their interaction with the ISM, highlighted in \citet{ferrara23}, are relevant only when the ISM is dense \add{($\gtrsim 10^5$ $m_\mathrm{p}$ cm$^{-3}$; \hbox{\citealt{Terlevich1992}})}, as it is in the first free-fall times of the first starbursts, prior to the onset of effective SN feedback.  Soon after, the ISM becomes dilute (in gas density,  metals and dust) due to both the high SFE of conversion of gas into stars and the super-Eddington winds.
The SN-driven steady wind (\se{wind}) keeps the ISM dilute for the rest of the $\sim\!100\Myr$ global FFB phase (\se{gas_frac}).
In such a dilute ISM, the radiation-driven wind is expected to become negligible, as there is only little dust to respond to the radiation pressure.    
In turn, the dilute ISM allows the SN-driven wind stream freely and be as effective as it could.
In more detail, 
there could be repeating episodes of a dense ISM (dusty and red) followed by a dilute ISM (dust-free and blue)
in a characteristic duty cycle, as eluded to in \citet{ferrara23}. 
This is as predicted in our FFB scenario, \se{sfh} and \citetalias{dekel23}.
During the dense episodes, one may expect bursty super-Eddington ejection on top of the steady SN-driven wind, but typically only at the very high redshifts indicated by \equ{z_sSFR}. 

% conclude
We conclude that super-Eddington ejections are potentially valid only at very high redshifts beyond the current main observed range, and even when they are, they are not likely to overwhelm the steady supernova winds that we use as a basis for our estimates of gas fraction, metallicity and dust attenuation.
Nevertheless, based on the SN-driven winds, we do predict low dust attenuation at high redshifts, 
which becomes negligible at very high redshifts, where it is also predicted to be removed in episodes of super-Eddington winds.

% comparison to obs
While \citet{ferrara23} argues that being dust-free is enough to bring the observed luminosities and SFRs
to the level of the JWST observations, this is not the case in massive halos at cosmic dawn based on the tentative comparisons shown in figures above between JWST observations and the standard nonFFB models; for example, as represented by the UM empirical model \citepalias{behroozi19} naively extrapolated to high redshifts and the standard SAM \citep{yung23}, both with no dust correction. 
Being dust-poor may help, especially at very high redshifts, but the main contribution to the excess of bright galaxies and high SFR at cosmic dawn is likely to be the enhanced SFE because of FFB, helped by the associated bursty SFE, maybe by a somewhat top-heavy IMF, and possibly by UV from AGN.

%============================
\subsection{Choice of ``standard" UM model}
\label{sec:disc_UM}

A word of caution is called for concerning the use of the empirical UM model \citepalias{behroozi19}.
We used it to normalize the SFE at low redshifts and low masses away from the FFB regime.
Then, in certain occasions, we compared our predictions for FFB with extrapolations of the UM model to
the relevant high redshifts.
One should be aware of the fact that the current version of the UM model has been derived based on
observations at lower redshifts, and the results with which we were comparing are based on
uncertain extrapolations to $z \ssim 10$.
These ``predictions'' of the UM at high redshifts should therefore be considered with a grain of salt.
This uncertainty in the UM model does not affect our predictions in the FFB regime, in which the UM predictions are not involved.
The UM can eventually be revised to refer to $z \ssim 10$ by fitting the model to new JWST observations
and thus recover the excess of bright galaxies at high redshifts.

%=============================
\subsection{FFB at lower redshifts}
\label{sec:low_z}

In principle, the FFB phenomenon may occur and be observable at redshifts and/or masses below the 
FFB threshold of \equ{thresholds}, with decreasing abundance further below the threshold line. 
These FFBs can naturally arise in our simple FFB model from the scatter in the 
parameters that control the 
density in the star-forming regions, which, once in the tails of their distributions, may bring the density to above the required threshold for FFB. For example, this may occur in disks of particularly small sizes that derive from small values of $\lambda$, or where the gas inflows are denser and/or more supersonic than the average for the given halo mass.
Alternatively, different mechanisms, such as wet galaxy mergers, intense instreaming, or other compaction events, 
may drive the required high densities and generate starbursts with high SFEs even at later times
\citep{Krumholz2012,zolotov15,turner15,lee16,usero2015,fisher22,querejeta19,lapiner23}.
These later high-SFE starbursts, however, may not necessarily follow the ``standard" FFB scenario 
worked out above and may therefore not share all the properties that were predicted for high-$z$ FFB galaxies.

%Difference from GMC
Densities above the FFB threshold of $n_{\rm crit} \ssim 3\stimes 10^3\cmc$ apparently appear in dense regions within giant molecular clouds (GMCs) in low-redshift disk galaxies. However, contrary to the FFB clouds, the GMCs are believed to have low overall SFEs that are likely suppressed by feedback, which actually tends to disrupt the clouds in a few dynamical times \citep[e.g.,][]{Krumholz17}.
Two obvious distinguishing features between GMCs and high-$z$ FFBs are the metallicity and the background 
disk density. The metallicity is near solar in GMCs while it is well below solar in FFBs, 
and the background disk density is much lower than the FFB density in GMCs while it is not far below 
the FFB density in FFBs.
In GMCs, the low disk density makes the cloud contraction (either at the original cloud formation or at its
reformation after disruption) be slow, on the disk free-fall timescale, 
which is longer than the $\sim\!1\Myr$ time window for FFB. 
During this contraction, the high metallicity makes $\tcool \slt \tff$ already in the early phases of cloud 
contraction, allowing star formation to occur before the cloud develops the dense core, thus leaving no time 
window free of supernova feedback. 
In FFBs, on the other hand, the high disk density makes the collapse rapid, on a cloud free-fall timescale 
of $\sim\!1\Myr$, and the low metallicity makes the cooling time drop below the free fall time and allow 
star formation only once the density exceeds the critical FFB density. This permits a burst free of feedback.

Furthermore, based on measurements of HCN/CO ratio, the regions in GMCs where the density is above 
$n_{\rm crit}$, which tend to be filamentary structures, are estimated to occupy only 
$\sim\!10^{-3}$ of the GMC volume in local disks, with a clear trend upwards with larger 
galactic-averaged density (Mark Krumholz, private communication; based on, e.g., \citealt{jimenez19}).
This makes the surface density in local GMCs of size $<100\pc$ lower than the critical surface density of 
$\Sigma_{\rm crit} \ssim 3000\Msun\pc^{-2}$ required for preventing significant gas ejection by stellar 
radiation pressure \citep{menon23}, but probably not so in high-redshift FFB clouds that reside in much 
denser background disks. Therefore, the SFR in local GMCs is not expected to be free of radiative feedback the way the high-$z$ clouds are.

%%%%%%%%%%%%%%%%%%%%%%%%%%%%%%%%% 10
\section{Conclusion}
\label{sec:conc}

\add{
% n and Sigma
As proposed analytically by \citetalias{dekel23},
feedback-free starbursts (FFB) are expected in gas clouds where the gas density exceeds a critical value of $n \ssim 3\stimes 10^3\cmc$. This ensures 
that the free-fall time is $\leq\! 1\Myr$, which is shorter than the timescale for feedback from supernovae and stellar winds, and also ensures that the cooling time is shorter than the free-fall time, allowing a starburst to occur on a free-fall timescale.
Moreover, the gas surface density in the FFB clouds should be above 
$\Sigma \ssim 2\stimes 10^3 \Msun\pc^{-2}$ to guarantee that radiative feedback is ineffective. 
}

\add{
% threshold mass and z
Consequently, enhanced SFE, of namely $\sfe \ssim 0.2 \sdash 1$, is expected in halos above a threshold in mass and redshift, roughly defined by the curve $\Mveight \ssim (1+z)_{10}^{-6.2}$, leading to distinguishable observable predictions, as follows.
}

\no\bul % SFR history
The average cosmological SFR density should be enhanced compared to the ``standard'' low-SFE scenario ---a small overdensity at $z \ssim 8$ and a large overdensity at $z \ssim 12$.

\no\bul % luminosity function
The galaxy mass and luminosity functions are expected to be correspondingly enhanced 
at $z \sgeq 8$, and gradually more so at higher masses and brighter magnitudes. 
For example, at $z \ssim 12$, galaxies of UV magnitudes of $-20$ are expected to be overabundant by more than an order of magnitude compared to the extrapolated standard empirical model of low SFE \citepalias{behroozi19}.

\no\bul % SFH
The SFH, during the global galaxy formation time of $\sim 100\Myr$, is expected to be bursty on a timescale of $\sim\! 10 \Myr$. In each such generation, a peak in SFR is expected to be followed by a low-SFR phase. 
This duty cycle should be translated to a global average SFE that is smaller than unity, possibly as small as 0.2. %
The predicted bursty SFH should be detectable by the distribution of the ratio of H$\alpha$ 
(or other Balmer lines) % added 
to UV SFR indicators. This burstiness should also contribute to the enhanced observed luminosities at the bright end.  

\no\bul % sizes
The FFB galaxies are expected to be compact, with typical stellar radii of 
$\Re \ssim 0.3\kpc\, (1+z)_{10}^{-2.5}$.

\no\bul % wind
The gas in a galaxy during its FFB phase is expected to form a steady wind driven by supernovae 
from post-FFB star clusters of earlier generations. 
The wind velocity is expected to be $V_{\rm w} \ssim 3,300\kms$ for $\eps \seq 1$ and $\sim\!700\kms$ 
for $\eps \seq 0.2$, observable through broad line wings of ${\rm FWHM}\ssim 2 V_{\rm w}$, 
which could be confused with AGN-driven winds.
The supernova-driven winds could possibly show two peaks centered on zero velocity, as well as low metallicity
and low ionization in BPT-like diagrams.

\no\bul
Using the steady density profile associated with the wind, we predict 
(a) a low gas fraction of approximately $\fg \sleq 0.05\,(1+z)_{10}^{-1}$,
(b) low metallicity in the FFB star-forming clouds of $Z \ssim 0.1\Zsun$,
and (c) low attenuation by dust with obscured fractions of $\sim\! 0.4$ and $0.2$ at $z\seq 9$ and $15,$ respectively. This diminished attenuation is also expected to contribute to the enhanced observed luminosity at high redshifts.

\add{%
In addition, it is worth reiterating several other observable predictions made by \citetalias{dekel23}.
}

\add{%
\no\bul % star clusters
A galaxy at the end of the FFB phase is predicted to be a compact assembly of thousands of star clusters, typically of $\sim\!10^6\Msun$ each.
These clusters form in gas clouds of ~$\sgeq 7\pc$ in radius.
}

\add{%
\no\bul
The feeding of the galaxy is expected to be through inflowing cold streams from the cosmic web, with no shock-heated gas in the halo, and minor outflows.   
}

Several galaxy properties are not significantly constrained by the FFB model. 
For example,
%\no\bul % morphology unconstrained
the global morphology could be of clumpy disks or clumpy spherical shells or an irregular morphology in between. 
%\no\bulo
As another example,
an ad hoc top-heavy IMF and/or UV from AGN may also contribute to the enhanced luminosities, but these factors are not necessary to explain the bright excess.

Ongoing and future studies of the FFB scenario and its implications include the following items: 

\no\bul
\add{
In \citet{Libanore2023}, we show that the FFB phase is expected to be detectable in the power spectrum of the 21cm signal, which will be observed by the Hydrogen Epoch of Reionization Array (HERA, \citealt{HERA}) at $z \sgeq 15$, 
but the net effect on the reionization history is expected to be negligible. 
}

\no\bul
\add{
The FFB star clusters could be the natural sites of massive black hole (BH) seeds due to core collapse that occurs in a few free-fall times, and is sped up by the broad stellar mass spectrum spectrum in the young clusters and the possible cluster rotation and flattening. While the stellar clusters disrupt each other rather quickly, these seeds, which populate compact galactic disks, merge into supermassive black holes with high BH-to-stellar mass ratios (Dekel et al. in prep.).
This post-FFB phase is yet to be properly investigated and simulated.
}

\no\bul
The FFB scenario and its observable predictions are to be studied using cosmological simulations in order to deepen, sharpen, and quantify our theoretical understanding of galaxy formation at cosmic dawn. 
This can be done via semi-analytic simulations in which FFB is incorporated by subgrid models.
Ultimately, it should be done using full hydro-gravitational cosmological simulations, which resolve the processes of star formation and feedback that should enable FFB.

In summary,
we obtained a large body of observable predictions, which should help us to distinguish the robust FFB scenario from other potential mechanisms that could contribute to the overabundance of bright galaxies at high redshifts.
Tentative, qualitative comparisons to early observational results from JWST seem to indicate qualitative agreement with the FFB predictions. These predictions should be quantitatively compared to upcoming observations by JWST.  

\medskip
% \section*{DATA AVAILABILITY}
To help others perform comparisons to other models, simulations, and
observations, 
we provide the digital tables and code to produce the predictions online.%
\footnote{\url{https://github.com/syrte/ffb_predict}}
Data and results underlying
this article will be shared upon reasonable request to the corresponding author.

\begin{acknowledgements}
We are grateful for stimulating discussions with
Josephine Baggen, Guilermo Barro, Peter Behroozi, Caitlin Casey, Katherine Chworowsky,
Andrea Ferrara, Steven Finkelstein, Yuichi Harikane, Thomas Harvey, Mark Krumholz, Elizabeth McGarth,  Shyam H.\ Menon, Yongzhong Qian,
Brant Robertson, Xuejian Shen, Rachel Somerville, Romain Teyssier, 
Andrea Weibel, Haojing Yan, Zhiyuan Yao, and Aaron Yung. 
ZL thanks Hengxiao Guo, Peng Wang, Jiaxin Han, and Rachel Somerville for hosting his visit at SHAO, SJTU, and CCA/Flatiron during the preparation of this work.
This work was supported by the Israel Science Foundation Grants
ISF 861/20 (ZL, AD) and 2190/20 (KCS).
ZL acknowledges the funding from the European Union’s Horizon 2020 research and innovation programme under the Marie Skłodowska-Curie grant 101109759 (``CuspCore'').
% 2190/20 (YB) and 3061/21 (NM, ZL),
%and by the DFG/DIP grant STE1869/2-1 GE625/17-1 (AD, KCS, ZL).
HA acknowledges support from the Zuckerman Postdoctoral Scholars Program.
\end{acknowledgements}


\bibliographystyle{mnras} % lizz
%%\interlinepenalty=10000 % to fix problem with certain references
\bibliography{z10}

\end{document}